\documentclass{aa}
\usepackage[varg]{txfonts}
\bibpunct{(}{)}{;}{a}{}{,}
\usepackage{tabularray}
\usepackage{footnote}
\usepackage{longtable}
\usepackage{lscape}
\usepackage{verbatim}
\usepackage{makecell}
\usepackage{multirow}
\usepackage{booktabs}
\usepackage{soul}
\usepackage{gensymb}
\usepackage{graphicx}
\usepackage{placeins}
\usepackage{ragged2e}
\usepackage{hyperref}
\usepackage{array}
\usepackage{subcaption}
\usepackage{textcomp}
\usepackage{float}
\usepackage{lastpage}
\usepackage{listings}

\newcommand\rurl[1]{%
  \href{http://#1}{\nolinkurl{#1}}%
}

\makeatletter
\renewcommand*\aa@pageof{, page \thepage{} of \pageref*{LastPage}}
\makeatother

\begin{document}

\title{Unsupervised learning for variability detection with \textit{Gaia} DR3 photometry}
\subtitle{The main sequence–white dwarf valley}

   \author{P. Ranaivomanana 
          \inst{1,2} \and C. Johnston \inst{3, 2} \and G. Iorio \inst{4} \and P.J. Groot \inst{1,5,6,7}  \and M. Uzundag \inst{2}   \and T. Kupfer \inst{8,9} \and C. Aerts \inst{1,2,10}
          } 
   \institute{
   Department of Astrophysics/IMAPP, Radboud University, P.O.Box 9010, 6500 GL Nijmegen, The Netherlands\\
   \email{princy.ranaivomanana@ru.nl}
   \and 
   Instituut voor Sterrenkunde, KU Leuven, Celestijnenlaan 200D, 3001 Leuven, Belgium \and
   Astrophysics group, Department of Physics, University of Surrey, Guildford, GU2 7XH, United Kingdom \and Departament de Física Quàntica i Astrofísica, Institut de Ciències del Cosmos, Universitat de Barcelona, Martí i Franquès 1, E-08028 Barcelona, Spain \and 
   Department of Astronomy, University of Cape Town, Private Bag X3, Rondebosch, 7701, South Africa \and South African Astronomical Observatory, P.O. Box 9, Observatory, 7935, South Africa \and  The Inter-University Institute for Data Intensive Astronomy, University of Cape Town, Private Bag X3, Rondebosch, 7701, South Africa \and
   Hamburger Sternwarte, University of Hamburg, Gojenbergsweg 112, 21029 Hamburg, Germany \and 
   Texas Tech University, Department of Physics \& Astronomy, Box 41051, 79409, Lubbock, TX, USA \and
   Max Planck Institute for Astronomy, Königstuhl 17, 69117 Heidelberg, Germany 
    }

   \date{Received month day, year; accepted month day, year}

 
  \abstract
{
The unprecedented volume and quality of data from space- and ground-based telescopes present an opportunity for machine learning to identify new classes of variable stars and peculiar systems that may have been overlooked by traditional methods. The region between the main sequence and white-dwarf sequence in the colour–magnitude diagram (CMD) hosts a variety of astrophysically valuable and poorly characterised objects, including hot subdwarfs, pre-white dwarfs, and interacting binaries.
} 
{
Extending prior methodological work, this study investigates the potential of an unsupervised learning approach to scale effectively to larger stellar populations, including objects in crowded fields, and without the need for pre-selected catalogues, specifically focusing on 13\,405 sources selected from Gaia DR3 and lying in the selected region of the CMD.
   }
{
Our methodology incorporates unsupervised clustering techniques based primarily on statistical features extracted from {\it Gaia} DR3 epoch photometry. We used the t-distributed stochastic neighbour embedding (t-SNE) algorithm to identify variability classes, their subtypes, and spurious variability induced by instrumental effects. Feature importance was evaluated using SHapley Additive exPlanations (SHAP) values to identify the most influential parameters associated with each cluster.
}
{The clustering results revealed distinct groups, including hot subdwarfs, cataclysmic variables (CVs), eclipsing binaries, and objects in crowded fields, such as those in the Andromeda (M31) field. Several potential stellar subtypes also emerged within these clusters, such as pulsating hot subdwarfs exhibiting pure or hybrid (pressure and/or gravity) modes within the hot subdwarf cluster. Magnetic CVs and dwarf novae appeared in the CVs cluster. Feature evaluation further enabled the identification of a cluster dominated purely by photometric variability, as well as clusters associated with instrumental effects and crowded fields. Notably, objects previously labelled as RR Lyrae were found in an unexpected region of the CMD, potentially due to either unreliable astrometric measurements (e.g. due to binarity) or alternative evolutionary pathways.
}
{ This study emphasises the robustness of the proposed method in finding variable objects in a large region of the Gaia CMD, including variable hot subdwarfs and CVs, while demonstrating its efficiency in detecting variability in extended stellar populations. The proposed unsupervised learning framework demonstrates scalability to large datasets and yields promising results in identifying stellar subclasses.}


\titlerunning{Unsupervised learning for variability detection}
\authorrunning{Ranaivomanana et al.}
\maketitle
%
\section{Introduction}
The advent of large-scale time-domain surveys has revolutionised observational astronomy. Ground- and space-based surveys such as the Palomar Transient Factory (PTF; \citealt{Law2009}), the Zwicky Transient Facility (ZTF; \citealt{Bellm2019}), the {\it Gaia} mission \citep{GaiaCollab2023}, and the Transiting Exoplanet Survey Satellite (TESS; \citealt{Ricker2015}) have produced large volumes of high-cadence photometric and spectroscopic data. These datasets have enabled not only the discovery of new classes of astrophysical transients and variables, such as fast blue optical transients \citep{Drout2014} and blue large-amplitude pulsators (BLAPs; \citealt{Macfarlane2015,Pietrukowicz2017}), but also the robust statistical characterisation of previously under-represented or poorly understood stellar populations, including hot subdwarfs and pre-white dwarfs \citep{Heber2016,Geier2017,Eyer2023}, EL CVn systems \citep{Roestel2018}, and detached double white dwarf binaries \citep{Burdge2019,Burdge2020}. Additionally, recently developed and forthcoming facilities such as BlackGEM \citep{Groot2024}, the Vera Rubin Observatory’s Legacy Survey of Space and Time (VRO/LSST; \citealt{Ivezic2019}), and the PLAnetary Transits and Oscillations of Stars (PLATO; \citealt{Rauer2025}) mission will continue to produce large datasets and thereby increase the probability of discovering new classes of astronomical objects.

In order to efficiently extract scientifically meaningful patterns from these large datasets, the astronomical community has increasingly adopted machine learning (ML) and deep learning (DL) methods. These techniques have become particularly prominent in the automated detection, classification, and clustering of variable stars, supernovae, and other transient phenomena (e.g. \citealt{Bloom2012,Villar2020,Pantoja2022,Ranaivomanana2025}). Supervised learning methods have been widely used to classify known types of variability, often relying on labelled training sets constructed from light curve morphology or statistical parameters \citep{Debosscher2007,Blomme2011,Richards2011,Aguirre2019}. However, supervised methods are limited by the availability of these training datasets and may fail to identify novel or rare types of variability.

To address this limitation, unsupervised ML approaches, particularly dimensionality reduction and clustering algorithms, are used to reveal hidden structure or patterns, as well as peculiarities in the data, without relying on labelled training sets \citep{Vandermaaten2008,Jolliffe2016}. Among these, t-Distributed Stochastic Neighbour Embedding (t-SNE; \citealt{Vandermaaten2008}) and the uniform manifold approximation and projection (UMAP; \citealt{McInnes2018}) have proven powerful for visualising high-dimensional data in a lower-dimensional space, revealing latent structures and relationships that are not immediately obvious in raw data. In astronomy, both algorithms have been applied successfully in a variety of contexts, including gamma-ray burst classification \citep{Jespersen2020,Zhu2024}, finding white dwarfs' hidden companions \citep{Perez2025}, and classification of eclipsing binaries \citep{Kochoska2017}.

This work extends our previous study, in which we developed an unsupervised ML framework based on t-SNE for detecting photometric variability in hot subdwarfs observed with \textit{Gaia} DR3 multi-epoch photometry (\citealt{Ranaivomanana2025}, hereafter Paper I). In Paper I, our analysis was limited to 1576 objects pre-selected from a catalogue of hot subdwarfs compiled by \citet{Culpan2022}. In the present study, we broaden the scope to a more diverse stellar population located in the valley between the main sequence and the white dwarf cooling sequence in the colour-magnitude diagram (CMD). This region encompasses a wide variety of stellar types of interest to the understanding of binary evolutionary pathways, including hot subdwarfs, pre-white dwarfs, cataclysmic variables (CVs), and compact binaries, many of which exhibit variability patterns not easily captured by traditional classification methods. As a large fraction of the objects in this transitional region remain poorly studied, identifying and characterising additional sources is essential for understanding their variability and constraining their evolutionary pathways.

Building upon the work presented in Paper I, the primary aim of this study is to demonstrate that our unsupervised learning framework is scalable to larger stellar populations and that it can potentially recover and separate distinct populations across the region between the main sequence and the white-dwarf sequence, without relying on pre-selected catalogues. In contrast to Paper I, which analysed the pre-selected sample of 1576 hot-subdwarf candidates \citep{Culpan2022}, here we apply the same feature extraction, dimensionality reduction, and clustering techniques, but to a much broader sample of 13\,405 objects. This scalability test is important because it demonstrates the method’s robustness when applied to a larger and more diverse dataset.

Additionally, the focus here is on providing a general overview of variability across the dataset rather than analysing individual objects or assessing the completeness of classification catalogues, as was the main subject of Paper I. Particular emphasis is given to the evaluation of the performance of statistical features in characterising the identified clusters.

This paper delivers unsupervised ML classification of the variability of 
the objects between the main-sequence and the white dwarf sequence, while suggesting key statistical features for variability detection that can be generally applied to any photometric observations. In addition, the study highlights the impact of applying data quality cuts on variability classification. The structure of this paper is as follows: In Sect.~\ref{sec:data_and_method}, we describe the data and methods. The clustering results are presented in Sect.~\ref{sec:results}, while the analysis of data quality cuts is discussed in Sect.~\ref{sec:ruwe_cut}. Our conclusion and future prospects are provided in Sect.~\ref{sec:conclusion}

\section{Data and methods}\label{sec:data_and_method}
Data were collected using publicly available datasets from {\it Gaia} DR3 \citep{GaiaCollab2023}. The {\it Gaia} mission provides photometric data in three main bands: the broadband G (330-1050 nm), the blue passband BP (330-680 nm), and the red passband RP (640-1050 nm).
To prepare our data for ML analysis, we followed a structured workflow that integrates target selection, data extraction, and feature extraction. The following sections describe these steps. 
\begin{figure*}
    \centering
\includegraphics[width=0.92\linewidth]{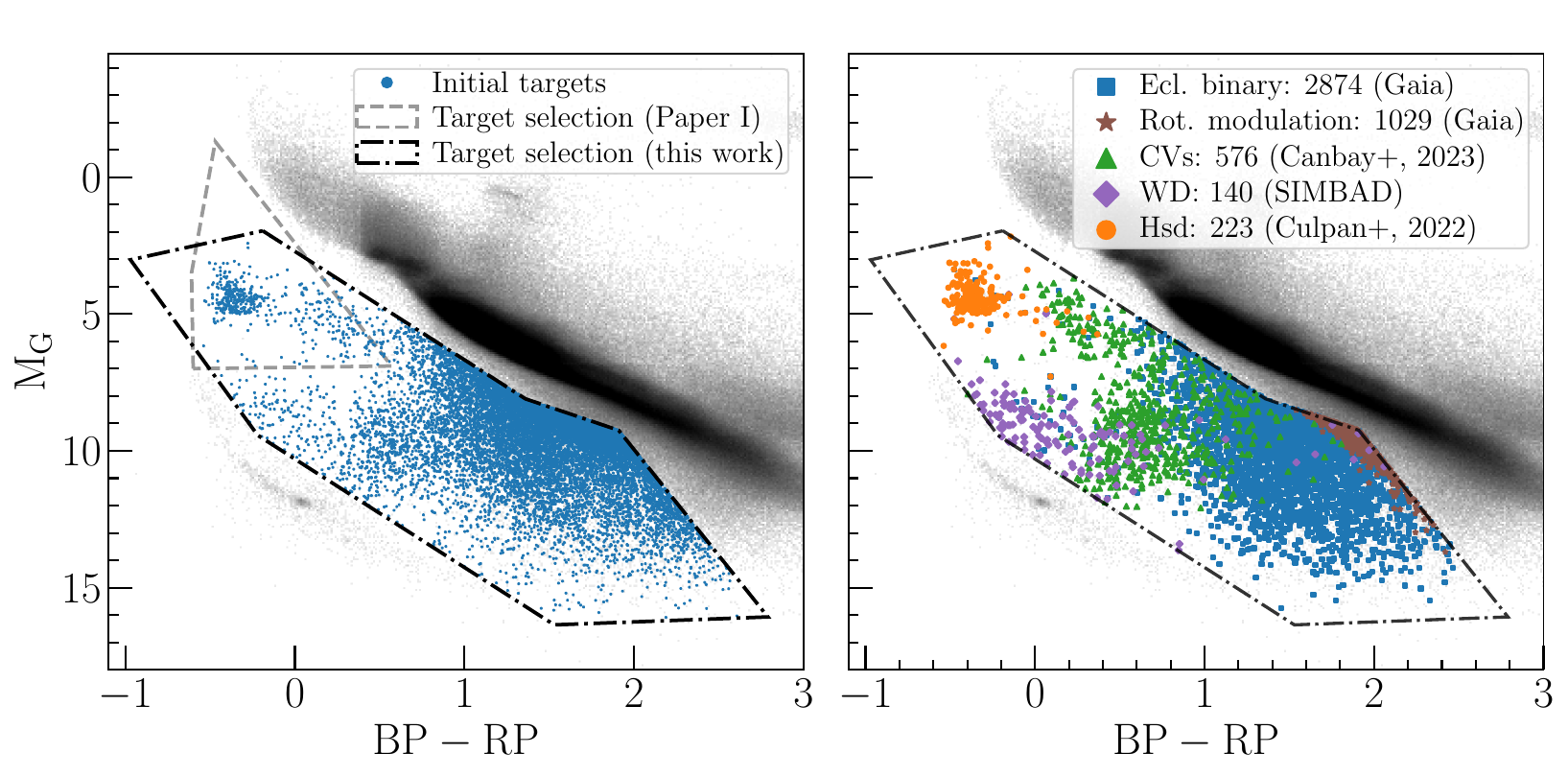}
\caption{\small Colour–magnitude diagrams, with grey background points representing all selected Gaia DR3 sources within 1 kpc. Left panel: blue points show the 18,085 initial targets drawn from the grey background sources within the black dash-dotted polygon. The dashed grey polygon marks the region from which the targets in Paper I were selected. Right panel: the identified stellar classes among the 13,405 final targets within the same black dash-dotted polygon, namely hot subdwarfs from Paper I (orange circles), eclipsing binaries from \textit{Gaia} classification (blue squares), solar-like rotational modulation stars from Gaia classification (brown stars), CVs from \cite{Canbay2023} catalogue (green triangles), white dwarfs from the SIMBAD database (purple diamonds), and hot subdwarfs from \citep{Culpan2022} catalogue. The dashed grey polygon indicates the freely selected target region.}
    \label{fig:cmd_known_obj}
\end{figure*}
\subsection{Target selection}
To extract the {\it Gaia} objects, we selected all sources within 1 kpc to mostly avoid Galactic extinction and reddening. We also required reliable parallax measurements (\texttt{parallax\_over\_error} $> 5$) and the availability of {\it Gaia} light curves (\texttt{has\_epoch\_photometry='True'}), with at least 25 observations in the {\it Gaia} G band (\texttt{num\_selected\_g\_fov} $> 24$), which we considered as the minimum necessary to detect photometric variability \citep{Ranaivomanana2025, Morales-Rueda2006}. These requirements were implemented in the {\it Gaia} Astronomical Data Query Language (ADQL) query form\footnote{\url{https://gea.esac.esa.int/archive/}}  when we ran the data extraction (see the Appendix for the full ADQL query). The query resulted in 2,080,613 objects, where distances in parsec (pc) were estimated by a simple 1/parallax estimation to compute the absolute G magnitudes $\rm M_G$. Using a more sophisticated method for distance determination \citep{Bailer-Jones2015} yielded very small differences due to the (pre-selected) high-quality parallax measurements. In the diagram, our initial sample was drawn from a region between the main-sequence and white-dwarf sequence, as indicated by the grey dashed line on the right panel of Fig.~\ref{fig:cmd_known_obj}. This was done by making a free selection in the area between the two sequences using TOPCAT\citep{Taylor2005}, while avoiding densely populated areas from both sequences\footnote{The TOPCAT expression for the area selection is: isInside(BP–RP, $\rm M_G$, –0.19, 1.96, 1.36, 8.10, 1.91, 9.25, 2.79, 16.07, 1.53, 16.36, –0.22, 9.43, –0.97, 3.02)}. Since these objects are further processed and classified by an ML algorithm, we could make a free selection in the CMD without the need to rely on traditional colour-selection criteria. As a result, we obtained 18\,085 objects between the main sequence and white dwarf sequence, as shown by the blue data points on the left panel of Fig.~\ref{fig:cmd_known_obj}. 

{\it Gaia}'s epoch photometry provides light curves for objects in the G, BP, and RP bands, with each transit corresponding to a $\sim50$ s broad G-band exposure, while BP and RP fluxes are obtained simultaneously from low-resolution prism spectrophotometry \citep{Hodgkin2021,Riello2021}. {\it Gaia} light curves in the three Gaia bands were extracted using the \texttt{astroquery.Gaia} Python package \citep{Ginsburg2019}. The value \texttt{EPOCH\_PHOTOMETRY} was specified for the \texttt{retrieval\_type} parameter in the package when extracting the light curves. Additionally, a light curve quality flag known as \texttt{reject\_by\_variability} \citep{Holl2018} was applied to each light curve to exclude epochs rejected by the {\it Gaia} variability pipeline. By extracting the light curves of the 18\,085 targets and after applying the quality flag to the light curves, we found 13\,405 {\it Gaia} light curves with more than 25 observations \citep{Morales-Rueda2006} in the {\it Gaia} G, BP, and RP bands. These light curves serve as our final dataset on which the feature extraction and clustering analysis of the {\it Gaia} epoch photometry were based. In the following sections, we preprocessed their {\it Gaia} light curves for feature extraction.

\subsection{Feature extractions}\label{sec:Gaia_photometry}
The first stage in the feature extraction involved running a frequency search algorithm on the 13\,405 targets to find the dominant frequency in each of the G-, BP-, and RP-band light curves. The frequency search algorithm described in \cite{Ranaivomanana2023,Ranaivomanana2025} was used in this work, with a frequency trial range from zero to 360 day$^{-1}$. In brief, the frequency search approach consists of computing the Lomb-Scargle periodogram (LSP, \citealt{Lomb1976,Scargle1982}) and the Lafler-Kinman statistic ($\Theta$, \citealt{Clarke2002,Lafler1965}), and determining the dominant frequency in the so-called $\Psi$-periodogram, defined as $2*\rm LSP/\Theta$. The next step was to extract statistical and photometric features from the $\Psi$-periodogram and the light curves. This was done by following the feature extraction steps described in \cite{Ranaivomanana2025}, from which a total of 54 features were obtained from the {\it Gaia} summary statistics table\footnote{\url{https://doi.org/10.17876/Gaia/dr.3/92}}, 6 parameters from the {\it Gaia} source database, and a set of 24 computed statistical features extracted from the actual light curves, resulting in a total of 84 light curve features. Since the number of observations in the G, BP, and RP bands ({\tt N\_G, N\_BP, N\_RP}) are already included in the {\it Gaia} summary statistics, we did not include them in this work. Thus, we obtained a set of 81 features as input data for the {\it Gaia} light-curve clustering.
\begin{figure*}
  \centering
\begin{subfigure}[b]{0.32\textwidth}
        \centering
\includegraphics[width=\textwidth]{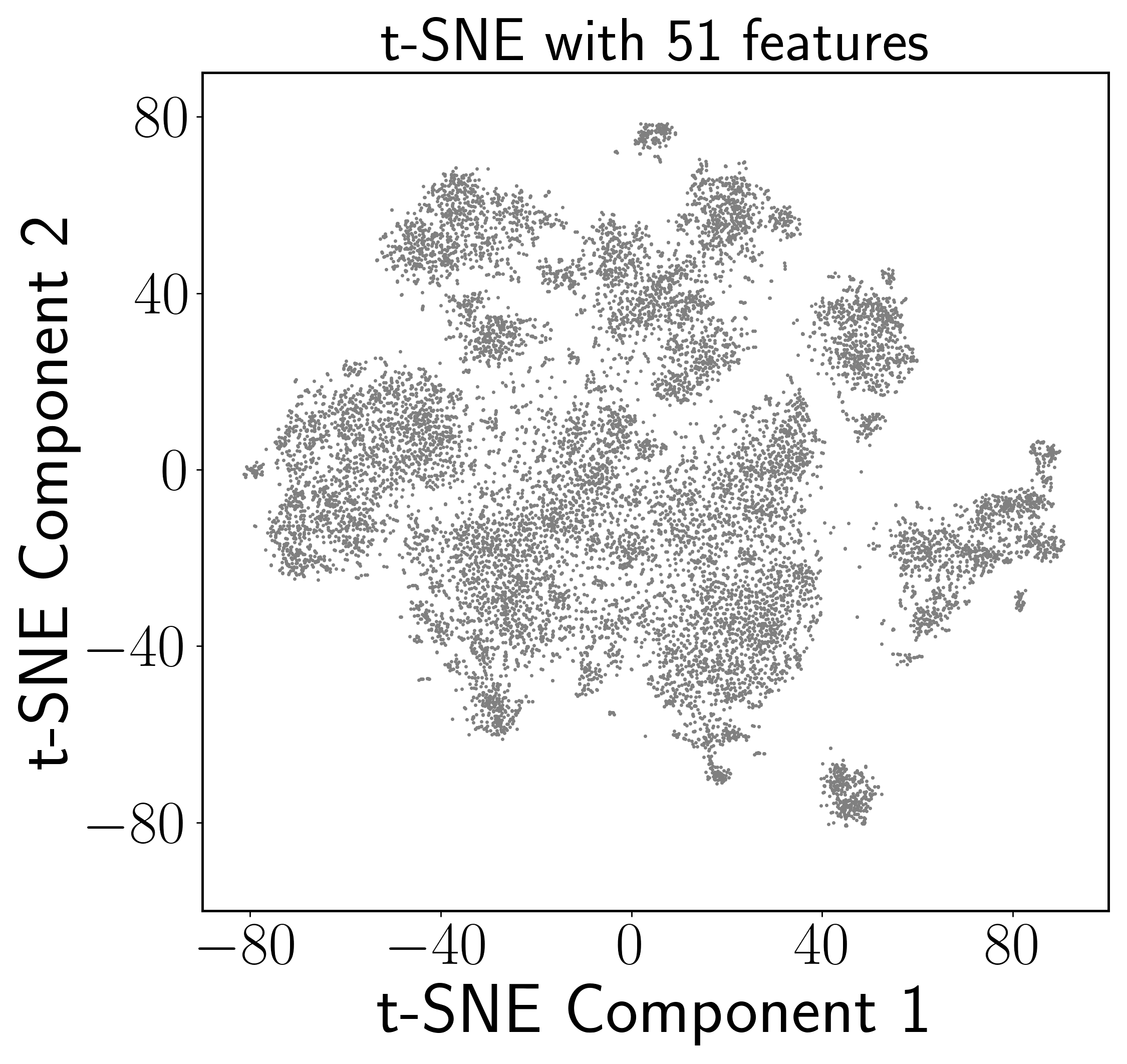}
        \caption{}\label{subfig:a}
    \end{subfigure}
    \hfill
    \begin{subfigure}[b]{0.32\textwidth}
        \centering
\includegraphics[width=\textwidth]{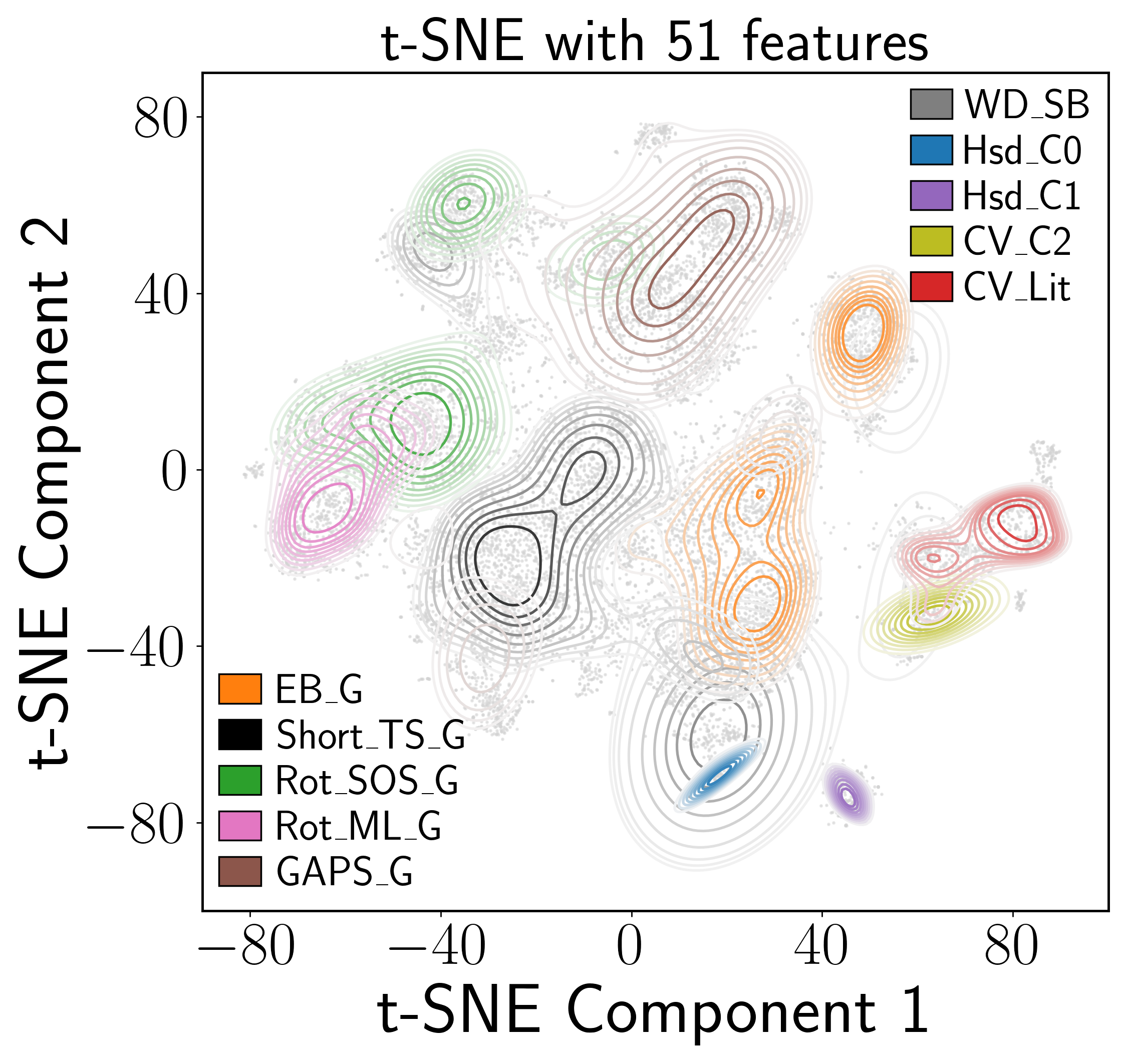}
        \caption{}\label{subfig:b}   
\end{subfigure}
\hfill
    \begin{subfigure}[b]{0.32\textwidth}
        \centering
\includegraphics[width=\textwidth]{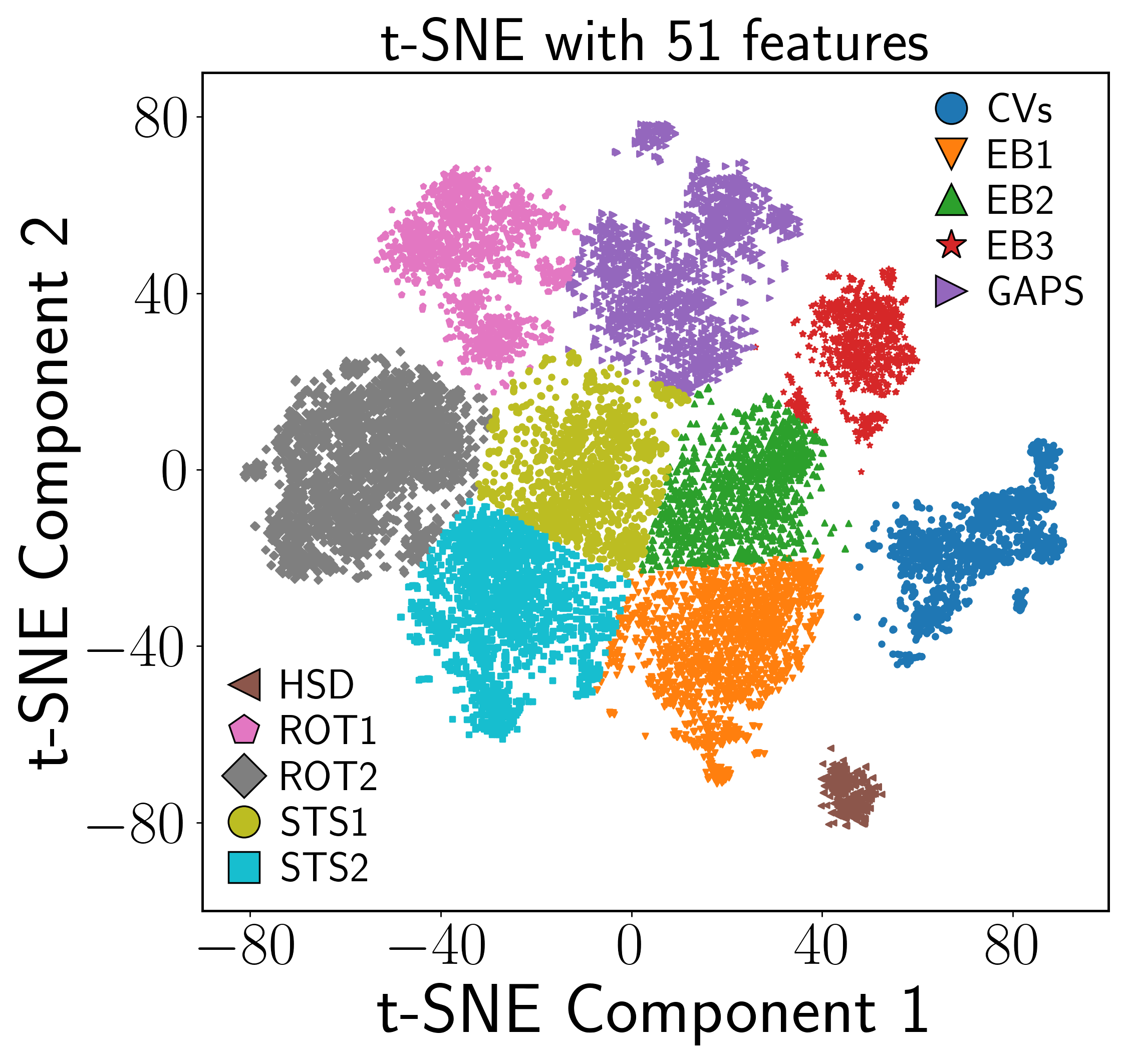}
        \caption{}\label{subfig:c}          
\end{subfigure}
\caption*{}
\begin{subfigure}[b]{0.32\textwidth}
        \centering
\includegraphics[width=\textwidth]{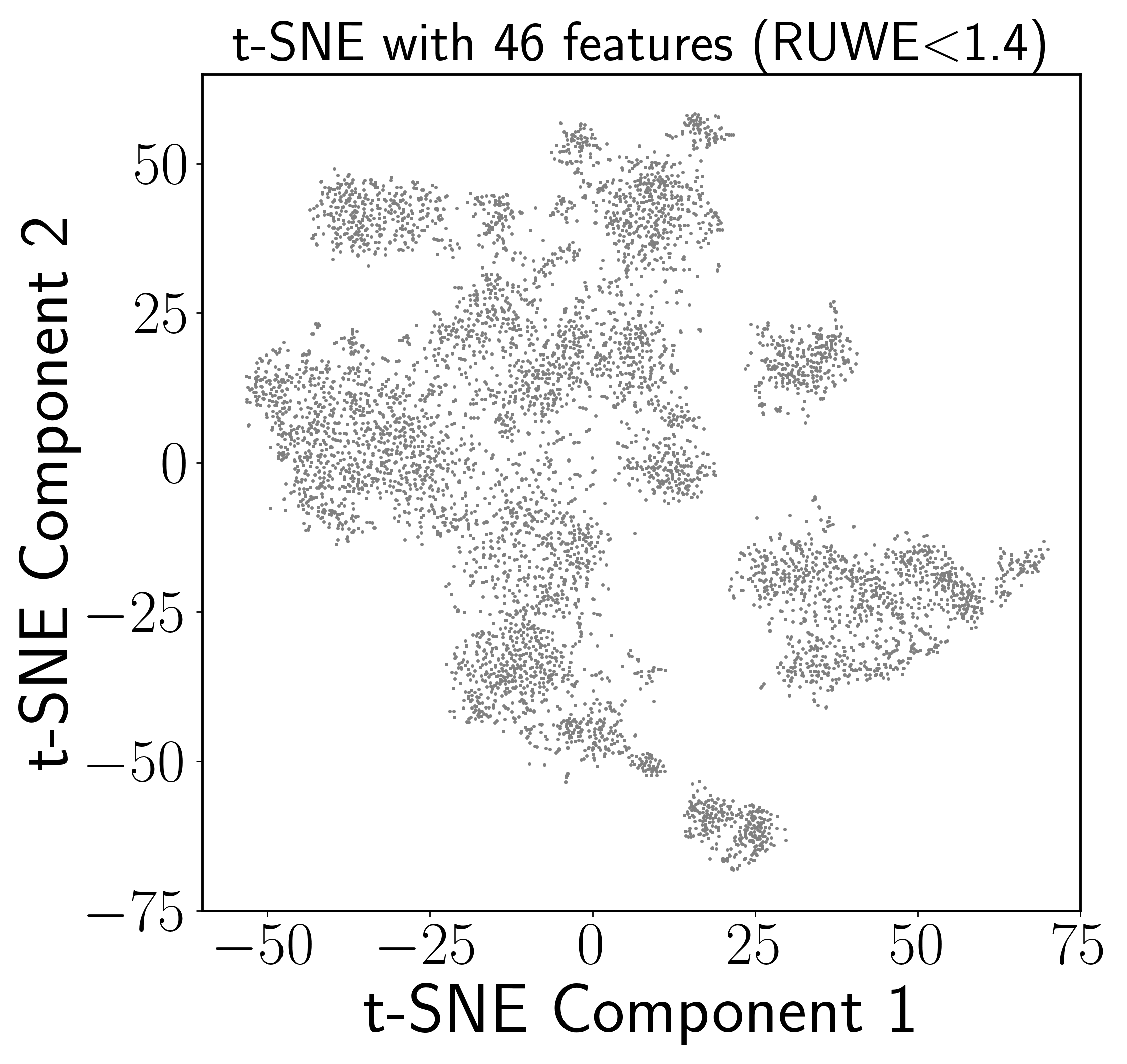}
        \caption{}\label{subfig:d}        
    \end{subfigure}
    \hfill
    \begin{subfigure}[b]{0.32\textwidth}
        \centering
\includegraphics[width=\textwidth]{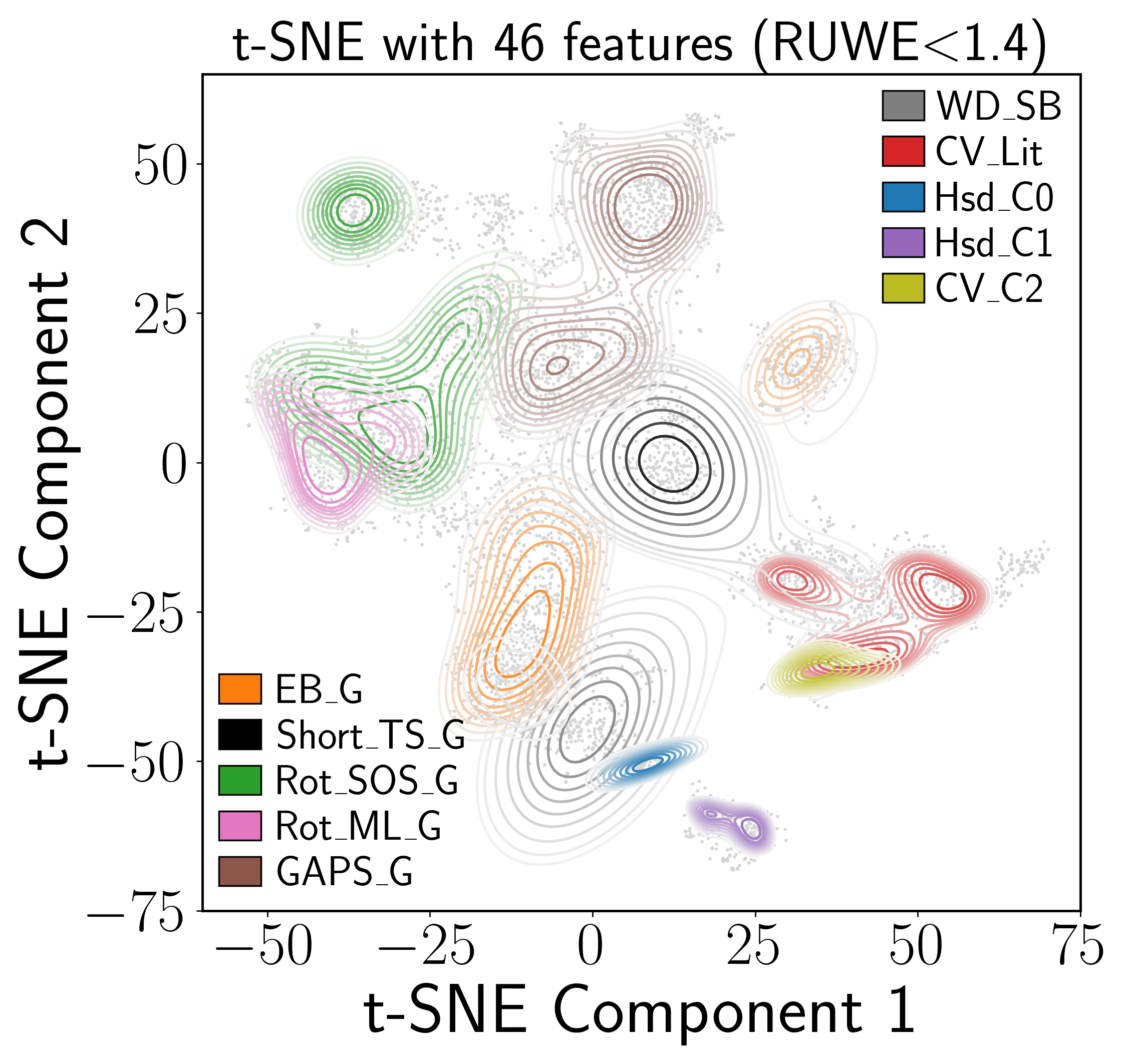}
        \caption{}\label{subfig:e}          
\end{subfigure}
\hfill
    \begin{subfigure}[b]{0.32\textwidth}
        \centering
\includegraphics[width=\textwidth]{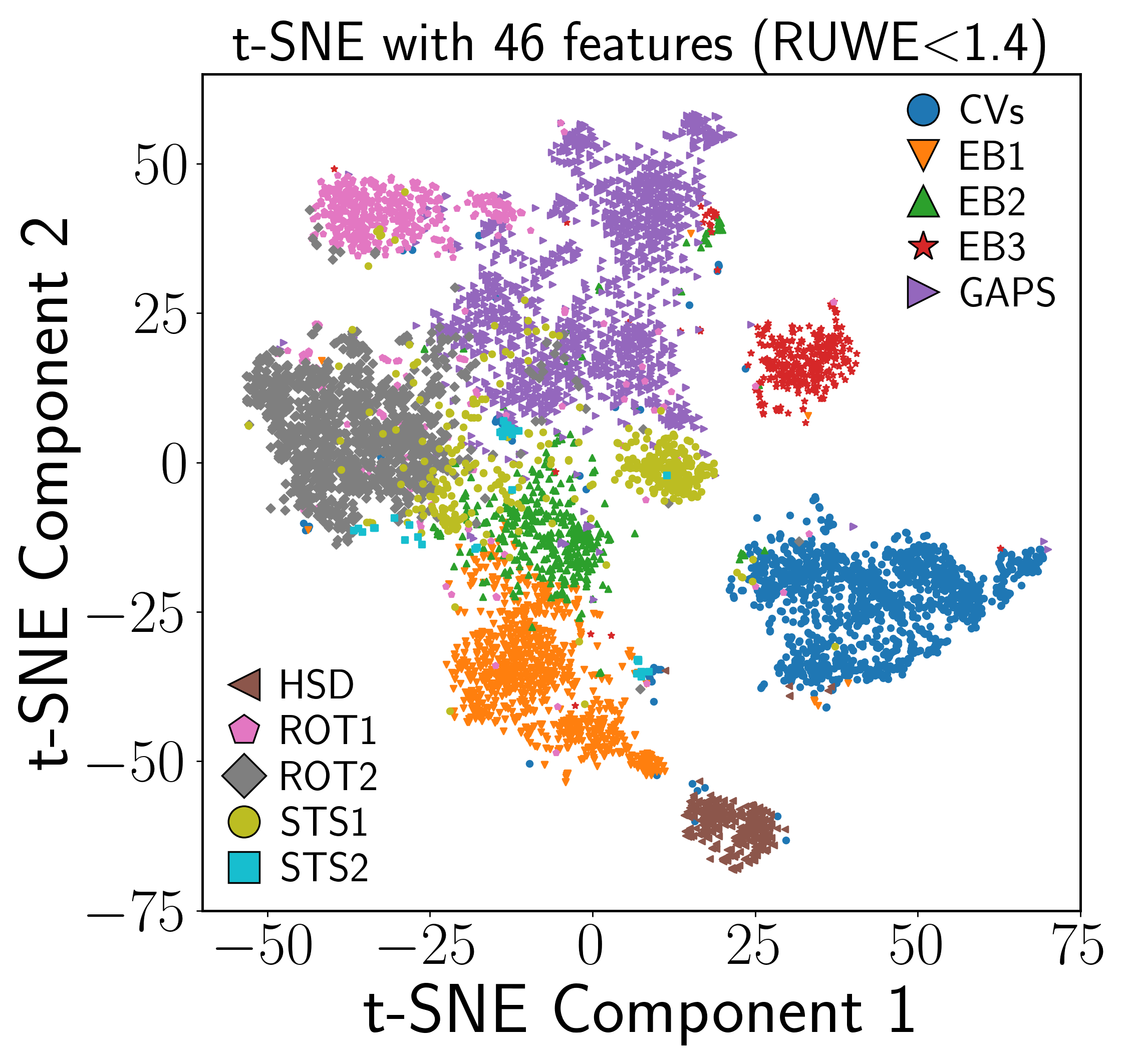}
        \caption{}\label{subfig:f}         
\end{subfigure}
\caption*{}
\caption{\small t-SNE embeddings for the original targets (a–c) and the reduced targets with RUWE$<$1.4 (d–f). Panels (b) and (e) show the t-SNE visualisations annotated with known classes from various sources: {\it Gaia} classifications (legends in the bottom left), SIMBAD (white dwarfs, labelled as WD\_SB), Paper I (Hsd\_C0, Hsd\_C1, CV\_C2), and cataclysmic variables from the literature (CV\_Lit). Panel (c) displays cluster labels derived from a Gaussian mixture model, where clusters are labelled according to known object types rather than numerical identifiers. Panel (f) shows the same cluster labels in panel (c) for the reduced dataset. The SOS and ML annotations in the legends refer to objects classified from the {\it Gaia} SOS and ML pipelines, respectively (see also Fig.~\ref{tab:nobj_per_cluster}).}\label{fig:tsne_51_feat}
\end{figure*}

\begin{figure*}
    \centering
    \begin{tabular}{c}
\includegraphics[width=0.95\linewidth]{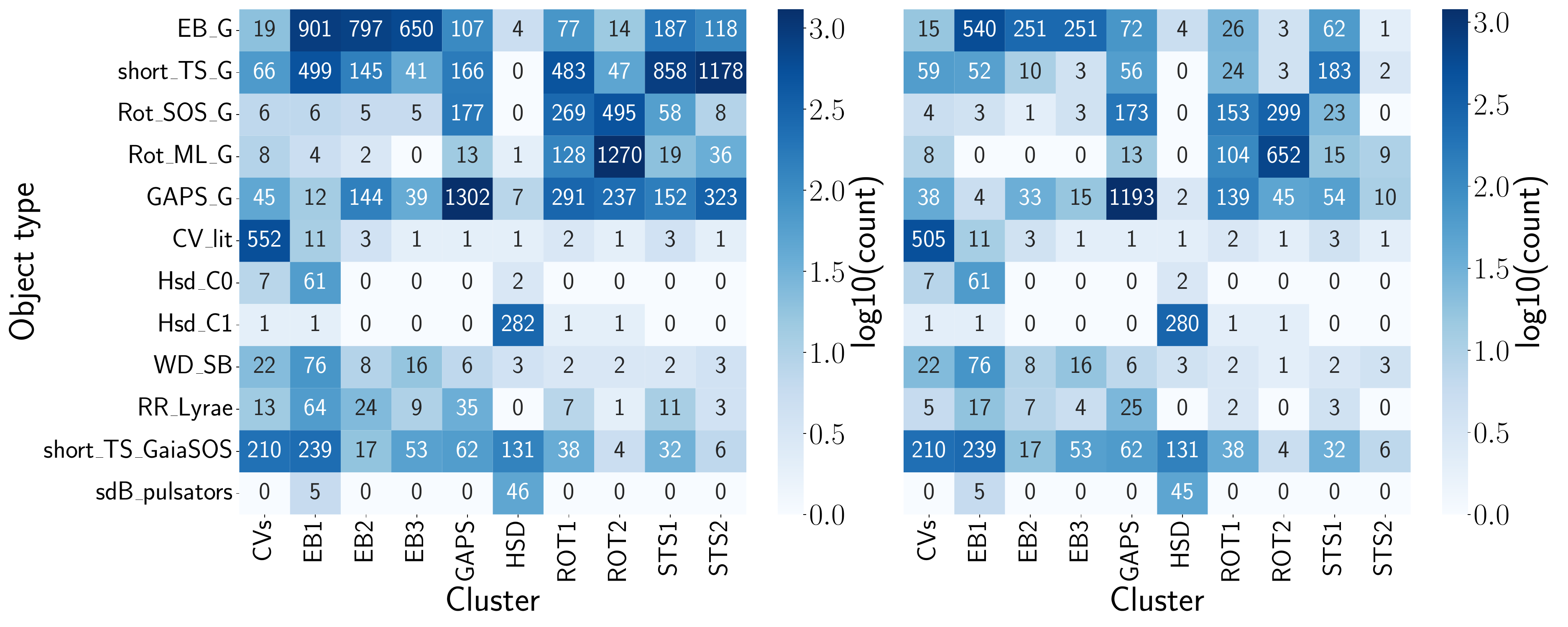}
\end{tabular}\caption{\small Number of known objects per cluster without a RUWE cut (left) and with the RUWE$<$1.4 cut applied (right). The x-axis (Cluster) shows the clusters defined in Fig.~\ref{subfig:c} and Fig.~\ref{subfig:f}, while the y-axis indicates the object types found in each cluster, as described in Table~\ref{tab:obj_definition}.}
    \label{tab:nobj_per_cluster}
\end{figure*}
After the features were extracted from the epoch photometry, a dimensionality reduction algorithm was applied to visualise these features in a 2D feature space and to use domain knowledge to interpret and validate the clustering results. In this work, dimensionality reduction was performed using the t-SNE algorithm as implemented in the openTSNE Python package \citep{Policar2021}. Compared to the original implementation \citep{Vandermaaten2008}, openTSNE offers several advantages in terms of scalability and transferability. More precisely, the openTSNE algorithm is computationally efficient over large datasets, and it also enables the embedding of new data into an existing t-SNE space. The latter is its unique feature compared to similar fast algorithms, such as the fast Fourier transform (FFT)-accelerated interpolation-based t-SNE (FIt-SNE) algorithm \citep{Linderman2019}.
\subsection{t-SNE optimisation and clustering}\label{sec:tsne_optimisation}
Following the steps outlined in Paper I and summarised in Fig.~\ref{fig:tsne_flowchart}, feature pairs with Pearson correlation coefficients greater than 0.95 were considered highly correlated. One feature from each pair was removed, resulting in a final set of 66 features. These features were then normalised to have zero mean and unit standard deviation (z-score normalisation) before optimising the t-SNE hyperparameters, namely perplexity and learning rate. The perplexity parameter reflects the effective number of local neighbours considered during similarity computations in t-SNE, while the learning rate determines the step size used in minimising the t-SNE cost function (see \citealt{Vandermaaten2008} for more details). The learning rate was fixed to \texttt{"auto"} while determining the optimal perplexity, which was varied from 30 to 100 in steps of 5. For each perplexity value, a Gaussian mixture model with 10 components (\texttt{n\_components=10}), reflecting the number of identified classes and sub-classes in Sect.~\ref{sec:results}, was used to cluster the resulting t-SNE embeddings. Given the smooth overlaps in the t-SNE embedding, GMM proved to be the most suitable choice: it explicitly models overlapping distributions and provides soft membership probabilities, which are essential when clusters overlap in feature space. Compared to the Density-Based Spatial Clustering of Applications with Noise (DBSCAN, \citealt{Ester1996}) algorithm that has been applied in similar contexts \citep[e.g.][]{Kochoska2017}, GMM produced more stable and interpretable cluster boundaries and is therefore the more appropriate method for this work.

Clustering performance was evaluated using the silhouette score \citep{Rousseeuw1987}, which evaluates clustering quality by measuring how well each data point fits within its assigned cluster compared to other clusters. As a result, a perplexity value of 70 yielded the highest silhouette score. Regarding the learning rate, setting it to \texttt{"auto"} produced the highest score compared to other tested values (ranging from 50 to 1000 in steps of 50). Cluster labels from the Gaussian mixture models were used to compute feature importance scores via a random forest model. To enhance clustering performance, the 66 features were ranked based on their importance scores. Using the optimised perplexity and learning rate values, as well as the ranked features, t-SNE was applied using the top 25 to 65 features. The number of features that produced the highest silhouette score was selected to generate the final clustering result shown in Fig.~\ref{fig:tsne_51_feat} (a–c), where 51 features were used. Using 5-fold cross-validation, the random forest classifier achieved an average accuracy of $0.89 \pm 0.01$, indicating it captured meaningful patterns. The resulting feature importance scores (Fig.~\ref{fig:feature_importance_score}) thus provide a reliable estimate of each feature’s contribution.

\section{Results}\label{sec:results}
To gain a general understanding of what each cluster represents, the 13,505 targets were cross-matched with catalogues of known objects built in Paper I, including hot subdwarfs \citep{Culpan2022, Ranaivomanana2025}, CVs \citep{Canbay2023}, and objects listed in the SIMBAD database \citep{Ochsenbein2000}. Thus, 223 known hot subdwarfs and 576 known CVs were identified in addition to 140 white dwarfs from SIMBAD. Note that amongst the hot subdwarfs and CVs were objects identified in Paper I referred to as cluster 0 (Hsd\_C0, 70 objects) and cluster 1 (Hsd\_C1, 286 objects) for candidate and known hot subdwarfs, and cluster 2 (CV\_C2, 98 objects) for CVs. As a reminder from Paper I, objects in cluster 0 exhibit clear periodic variability, whereas those in cluster 1 show weak or unclear variability patterns.
Since the targets in this work were limited to objects within 1 kpc, only a subset matched those identified in Paper I. 

Furthermore, {\it Gaia} DR3 provides variability classifications for approximately 9 million variable sources produced by ML classifiers \citep{Rimoldini2023}. The resulting classifications are followed by a dedicated pipeline known as Specific Object Studies (SOS) to validate individual classes, except for a few SOS pipelines, such as the SOS module for solar-like rotation modulation stars \citep{Distefano2023} and short-timescale (period $<$ 1 d) variables \citep{Roelens2018}, with candidate selections independent of the ML results \citep{Rimoldini2023}. Using the classifications published by these pipelines, we identified in our sample objects that were previously labelled, including 167 RR Lyrae stars \citep{Clementini2023}, 2,874 eclipsing binaries \citep{Mowlavi2023}, 792 short-timescale variables \citep{Rimoldini2022}, 2552 objects in the {\it Gaia} Andromeda Photometric survey (GAPS, \citealt{Evans2023}), and 1,029 and 1481 solar-like rotation modulation stars from the {\it Gaia} SOS pipeline \citep{Distefano2023} and {\it Gaia} ML classification, respectively. 

A colour–magnitude diagram of the objects with known classifications is shown in Fig.\ref{fig:cmd_known_obj}. Eclipsing binaries occupy the region between the main sequence and the white dwarf sequence, while validated rotational modulation stars from the {\it Gaia} SOS pipeline are located near the main sequence, at the boundary of the target selection. Note that the rotational modulation candidates were selected from the main-sequence region of the CMD using strict selection criteria (see Fig. 1 in \citealt{Distefano2023}).
\begin{figure}
    \centering
    \begin{tabular}{cc}
    \includegraphics[width=0.95\linewidth]{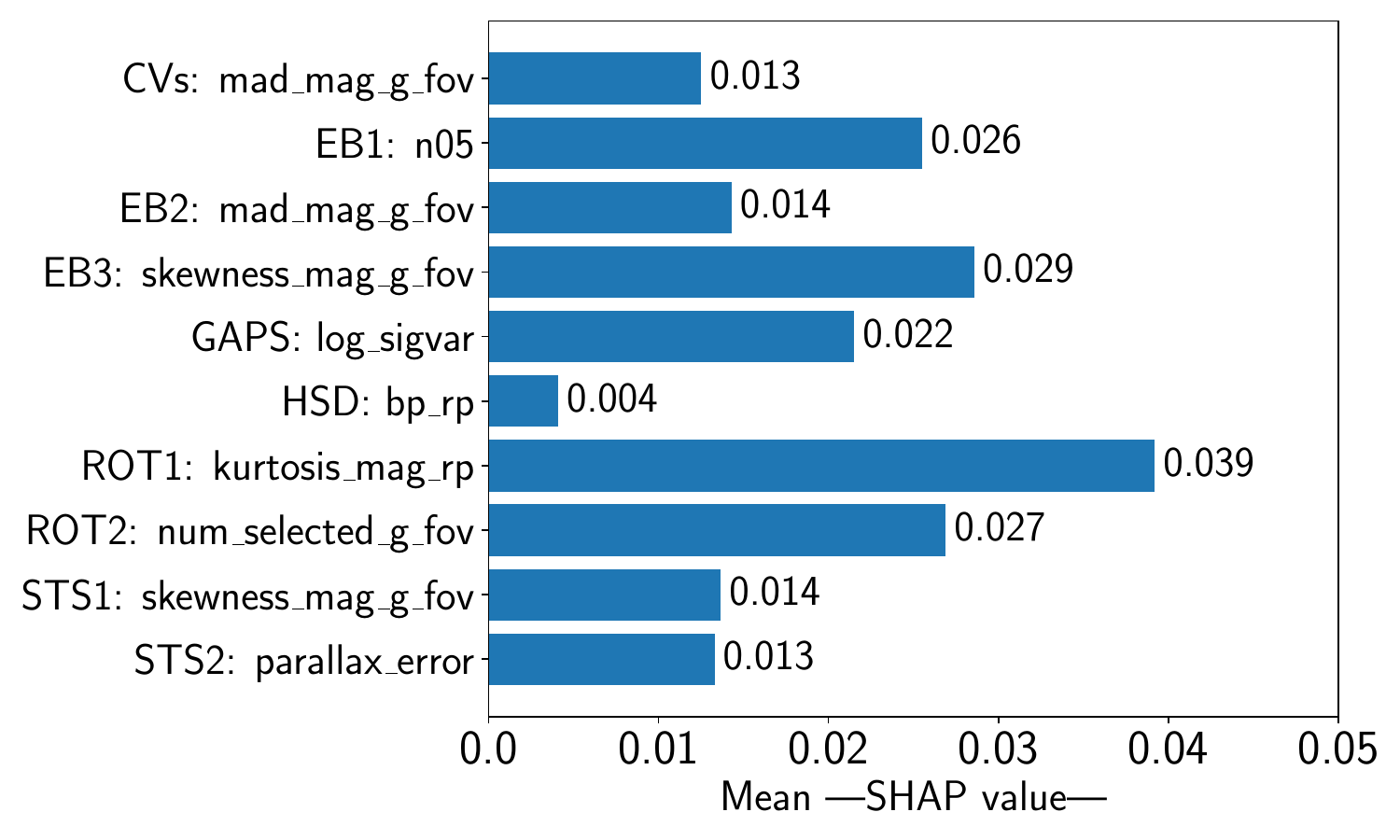} \\
    \includegraphics[width=0.95\linewidth]{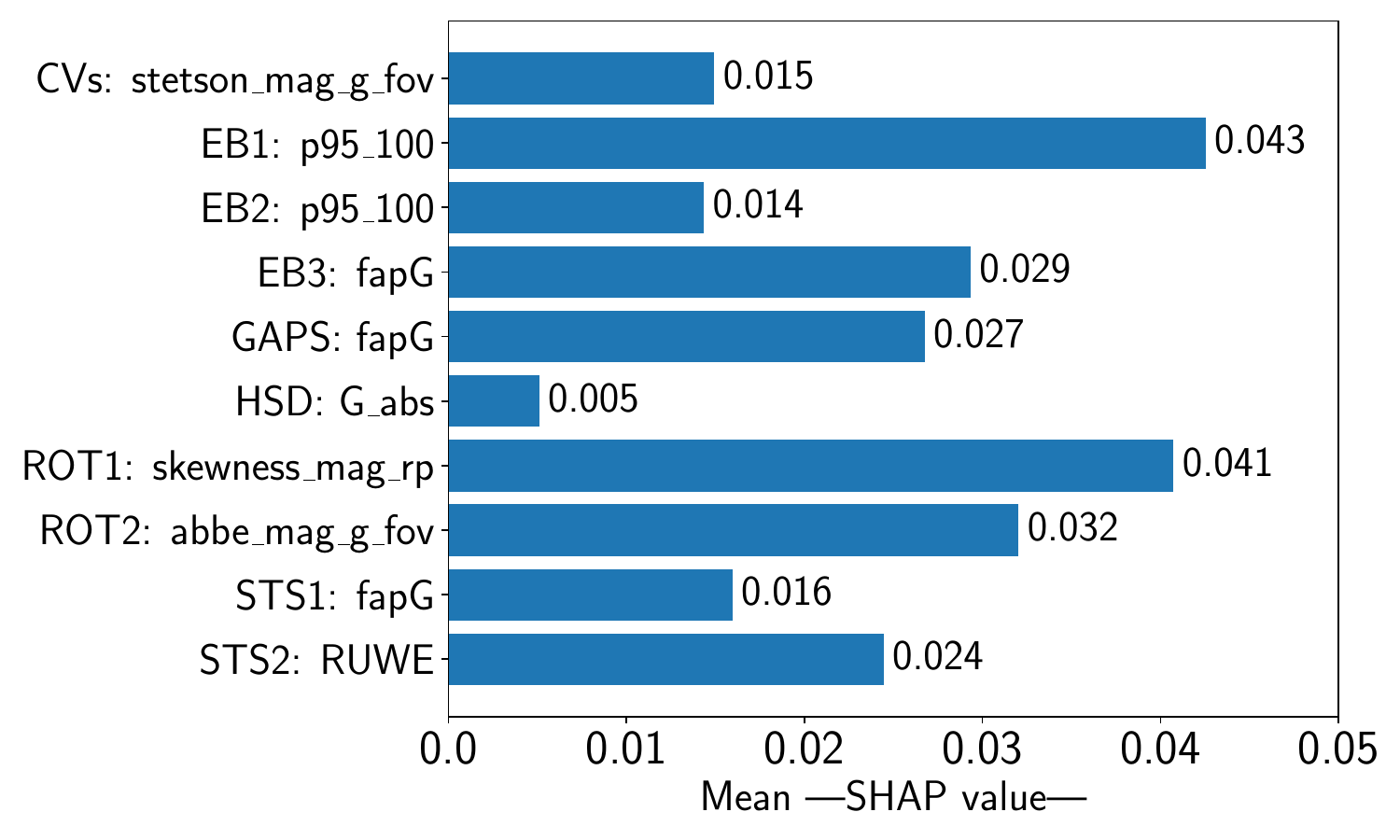}
    \end{tabular}\caption{\small SHapley Additive exPlanations (SHAP) values for the most important features in predicting each cluster: the top panel shows the highest-ranked feature, and the bottom panel shows the second-most important. SHAP values are expressed in log-odds units.}
    \label{fig:shap_values}
\end{figure}
\begin{figure}
    \centering
    \begin{tabular}{c}    \includegraphics[width=0.95\linewidth]{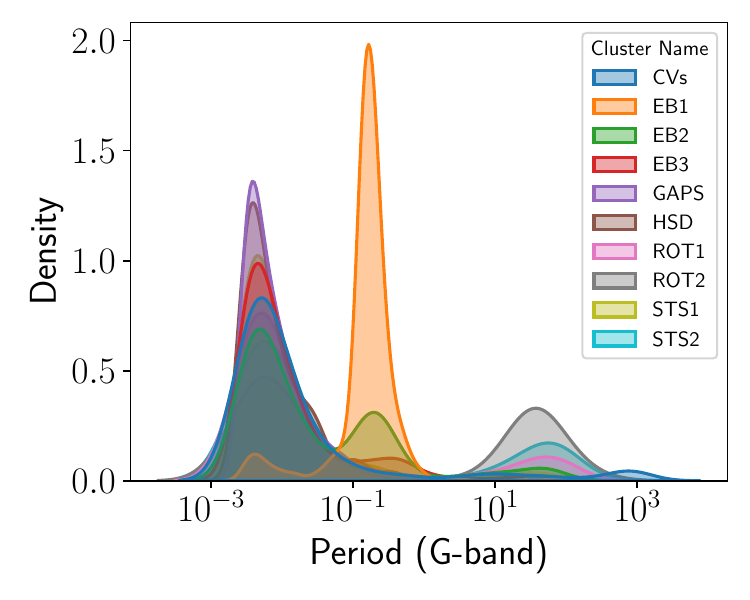}
    \end{tabular}\caption{\small {\it Gaia} G-band period distribution per cluster.}
    \label{fig:period_dist}
\end{figure}
\subsection{Dimensionality reduction implementation}\label{sec:clustering_analysis}
\subsection{t-SNE embeddings}
Figure~\ref{fig:tsne_51_feat} shows the resulting t-SNE embeddings. In sub-panels \ref{subfig:b} and \ref{subfig:e}, the locations of the above known classes are represented by density contour lines. These contours are drawn from Gaussian kernel density estimates using the \texttt{seaborn}\footnote{\url{https://seaborn.pydata.org/index.html}} Python package with the function \texttt{seaborn.kdeplot}. Note that the objects previously labelled as RR Lyrae stars are present everywhere in the t-SNE embeddings, particularly in the eclipsing binary clusters; therefore, they are not shown in Fig.~\ref{fig:tsne_51_feat} for a better visualisation. However, they are shown in Fig. \ref{fig:rrlyrae_tsne} in the Appendix and discussed further in Sect.~\ref{sec:rrlyrae}. The short-timescale variables overlap in the t-SNE embeddings with the clusters with the hot subdwarfs, white dwarfs, and CVs, and therefore they are not also shown in Fig.~\ref{fig:tsne_51_feat} for clarity purposes. The overlap is due to the fact that this class corresponds to objects with fast variability, defined as having periods less than 1 day \citep{Roelens2018}, which overlaps mostly with the range of periodicity in the aforementioned three classes. Apart from variables validated by the SOS pipeline, 3483 short-timescale variable candidates and 1481 solar-like rotation modulation stars from the {\it Gaia} machine-learning classification \citep{Rimoldini2023} were found in our sample. These objects are distributed somewhat distinctively in the t-SNE embeddings as shown in Fig.~\ref{fig:tsne_51_feat}, with a few overlaps with those classified from the SOS pipelines: 40 and 12 objects overlap for the short-timescale variables and rotation modulation stars, respectively.

The cross-matched sources allowed us to validate the clustering results shown in Fig.~\ref{fig:tsne_51_feat}, where each cluster generally represents a physically meaningful object class. For hot subdwarfs and CVs in particular, the results for Hsd\_C0, Hsd\_C1, and CV\_C2 are consistent with the findings in Paper I, where the three classes are distributed distinctively in the t-SNE embeddings. Of the 70 objects originally in Paper I’s Hsd\_C0 set that are present in our sample, 61 (87\%) lie in cluster EB1 in the current t-SNE embedding, with only 7 objects in the CV cluster and 2 in the HSD cluster. Conversely, of the 286 objects from Paper I’s Hsd\_C1 set, 282 (98.6\%) fall in the HSD cluster here. The Paper I CV candidate set (CV\_C2) likewise maps predominantly to the CVs cluster in this work. These mappings (Fig.~\ref{tab:nobj_per_cluster}) demonstrate that the three main clusters identified in Paper I remain distinct when the analysis is performed on a substantially larger and more diverse dataset (13,405 objects), confirming the stability of our unsupervised method. 

\subsection{Feature evaluation}
Now that each cluster in the t-SNE embeddings has been identified, it is important to examine which features contribute to assigning an object to a particular cluster. This analysis is especially useful for understanding why objects of the same type may belong to two or more distinct clusters. To evaluate the contribution of each feature to each cluster, the same approach as in previous sections was followed, using a Gaussian mixture model to predict class labels for a specified number of clusters.

Since the number of identified classes is approximately 10, and some classes span multiple clusters, the Gaussian mixture model was fitted with 10 components ({\tt n\_components=10}). The resulting clustering is shown in Fig. \ref{subfig:c}, where the 10 clusters were renamed based on the predominant type of objects identified in each cluster (see Fig.~\ref{tab:nobj_per_cluster}), rather than using the default numeric labels (e.g., Cluster 0 or Cluster 1). For instance, the cluster containing known hot subdwarfs was renamed "hot subdwarfs (HSD)" instead of "Cluster 0". Additionally, object types that appear in multiple clusters (e.g. EB) were given additional labels, such as EB1 and EB2. The number of known objects in each cluster is summarised in Fig.~\ref{tab:nobj_per_cluster}, which highlights the most prevalent object types per cluster.  

The output labels from the Gaussian mixture model were used to fit a random forest model to estimate feature importance scores. Since the goal here is to obtain importance scores for each individual cluster, SHapley Additive exPlanations (SHAP) values \citep{Lundberg2017} were used to quantify the contribution of each feature to the random forest predictions. SHAP values measure how much each feature increases or decreases a prediction relative to the average prediction. A summary plot of the first and second most contributing features for each cluster is shown in Fig.~\ref{fig:shap_values}. The relevance of these features is further supported by kernel density plots in Fig.~\ref{fig:kde_plots_features}, stressing their distribution per cluster. To better understand the detected variability periods within each cluster, the period distributions are shown in Fig.~\ref{fig:period_dist}, revealing three main distributions centred on timescales of minutes, hours, and days in the {\it Gaia} G band. The majority of the clusters (8 out of 10) exhibit short-period distributions on timescales of minutes. While genuine short-timescale variability may be present in these clusters, a significant fraction could result from aliasing effects, as discussed in \cite{Roelens2018}. Similarly, the long-period distribution seen in Fig.~\ref{fig:period_dist} may largely be attributed to aliasing frequencies, such as the {\it Gaia} precession period at 62.97 days \citep{Lebzelter2023}. On the other hand, the narrow peak around a few hours primarily corresponds to genuine variables, including eclipsing binaries, as described in Sect.~\ref{sec:eclipsing_binaries}. 

We now focus on investigating feature importances for each object class, especially those that appear in more than one cluster, including eclipsing binaries, solar-like rotational modulation variables, and short-timescale variables. This analysis aims to help identify the distinguishing characteristics between these clusters.

\subsubsection{Eclipsing binaries}\label{sec:eclipsing_binaries}

The distribution of eclipsing binaries from {\it Gaia} classification is shown in Fig.~\ref{subfig:b}, which are labelled as EB1, EB2, and EB3 in Fig.~\ref{subfig:c}. The SHAP value outputs in Fig.~\ref{fig:shap_values} for these clusters indicate that the features \texttt{p95\_100} and {\tt n05} are highly important for predicting EB1. The feature \texttt{p95\_100} represents the 95th percentile of the 100 strongest power values in the periodogram, whereas \texttt{n05} denotes the number of frequencies whose power $\Psi$ exceeds 0.5 in the normalised periodogram. These features are critical for identifying light curves with clear variability, as demonstrated in Paper I. This is supported by visual inspection of objects in cluster EB1, where 1497 out of 1703 objects show unambiguous variability, mostly consisting of eclipsing binaries.

In contrast, cluster EB2 also contains clearly variable objects, with \texttt{p95\_100} and \texttt{mad\_mag\_g\_fov} being the most important features. However, there are only a few of them since EB2 is contaminated by objects with noisy periodograms. This is demonstrated by the number of peaks above 0.5 of the normalised periodogram ({\tt n05}), where the 10th and 90th percentiles of {\tt n05} for EB2 are 17 and 452, respectively, while these values are 2 and 40 for EB1, respectively. This suggests a poorly constrained variability for EB2. 

Finally, the false alarm probability ({\tt FAP}) contributes the most to the prediction of EB3, where more than 80\% of objects in EB3 have {\tt FAP} values above 0.6. The variability observed in EB3 is likely associated with aliasing frequencies, indicating less reliable or spurious variability signatures.

\subsubsection{Short-timescale variables}
This category contains two clusters, namely STS1 (1333 objects) and STS2 (1688 objects). Firstly, the prediction for cluster STS1 is mainly driven by the {\tt FAP} feature and skewness in the G band (\texttt{skewness\_mag\_g\_fov}). The SHAP values for the two parameters are approximately the same, as seen in Fig. \ref{fig:shap_values}, suggesting that they have a similar impact on the model prediction. Although the majority (80\%) of cluster STS1's {\tt FAP} values are below 0.1 with a median value of detected periods of 9 min, the {\tt FAP} values may not reflect the period significance of such high-frequency variables \citep{VanderPlas2018}. Visual inspection shows that the STS1 cluster contains mostly noisy periodograms, most likely due to the sparsity of the {\it Gaia} sampling. Further observations would be required to confirm the variability in STS1. Regarding the skewness parameter, about 75\% of the objects in STS1 have negative skewness, which may suggest that their variability is likely caused by flaring events if only a few bright events are captured among mostly quiescent observations. However, this could be a result of selection effects since short-timescale variable candidates described in \cite{Rimoldini2023} have a good balance between negative and positive skewness values, where candidates are selected in such a way that –1.4 $<$ \texttt{skewness\_mag\_g\_fov} $<4$. 

Objects in STS2 are characterised by high RUWE values, where the majority (90\%) of the objects have RUWE $>2.6$. Compared to the overall population, objects in STS2 have higher parallax error with a median of 0.42 mas, while the median value for all the objects is 0.23 mas (excluding STS2). These objects could present rapid variability candidates in crowded fields or merely unresolved binary systems. 

\subsubsection{Solar-like rotational modulation}

This class of objects is divided into two clusters: solar-like rotational modulation 1 and 2, referred to as ROT1 and ROT2, respectively, as shown in Fig.~\ref{fig:tsne_51_feat}. ROT1 exhibits a stronger negative skewness in the RP band compared to the G band, with 90\% of its members having negative skewness values. These objects exhibit occasional bright outliers in their RP band light curves, most likely due to instrumental artefacts, contributing to the more negatively skewed distribution. Similarly, the kurtosis pattern in the RP band for ROT1 may also result from the bright outliers. On the other hand, ROT2 is characterised by lower Abbe values, with \texttt{abbe\_mag\_g\_fov} centred around 0.5, and a higher number of observations in the G band, with a median of 71 observations compared to 45 for the full sample. The lower Abbe values in ROT2 could indicate light curves with trends, pulsations, or transient events \citep{Mowlavi2014,Roelens2018}. The increased {\it Gaia} sampling for ROT2 is likely a result of the {\it Gaia} scanning law \citep{Rimoldini2023}. Additionally, the Gaia SOS rotation modulation selection requires segmentation of long-term, densely sampled time-series data \citep{Distefano2016,Distefano2023}, which contain more observations than are typical for Gaia sources. This selection effect leads to an increased number of identified observations and may also influence the Abbe value.

\subsubsection{Hot subdwarfs}
The SHAP values for the hot subdwarf (HSD) cluster suggest that the {\it Gaia} G-band absolute magnitude and BP$-$RP colour are the primary features driving their classification. These two parameters are known to characterise hot subdwarfs in the colour–magnitude diagram, confirming the robustness of the SHAP value analysis in identifying the most relevant features for each class. Moreover, the HSD cluster is the least contaminated, containing the majority (46 out of 50) of known pulsating hot subdwarfs \citep{Uzundag2024}. This cluster includes promising candidates for identifying pulsating hot subdwarfs through multiple observational campaigns. The variability of all objects in the HSD cluster has been studied in detail by \cite{Ranaivomanana2025}, except for 10 objects not included in their hot subdwarf training set from \cite{Culpan2022}.

Furthermore, a close view of the t-SNE embedding for the HSD cluster reveals two sub-clusters in the left panel of Fig.\ref{fig:zoomedin_tsne}, where pulsating hot subdwarfs from the literature \citep{Baran2024, Krzesinski2022} have been identified. HSD sub-cluster 0 contains pure pressure (p) and gravity (g) mode pulsating hot subdwarfs, while sub-cluster 1 includes both p- and g-mode pulsators, as well as hybrid (p+g) mode pulsators and g-mode pulsators in binary systems. Since the number of objects with known pulsation modes in both sub clusters is not statistically significant, it is not yet conclusive whether these two sub clusters represent hybrid and pure pulsators, respectively. We therefore present these as promising indications that merit confirmation with larger samples or targeted spectroscopy, but we do not claim definitive subclass classification here.
Additionally, both sub-clusters contain objects with low photometric amplitude variations, with median values of 7 mmag and 8 mmag for sub-cluster 0 and sub-cluster 1, respectively. As demonstrated in \cite{Ranaivomanana2025}, these amplitudes are too small to allow detection of clear variability in {\it Gaia}.

\subsubsection{Cataclysmic variables}
\begin{figure*}
    \centering
    \begin{tabular}{cc}
\includegraphics[width=0.42\linewidth]{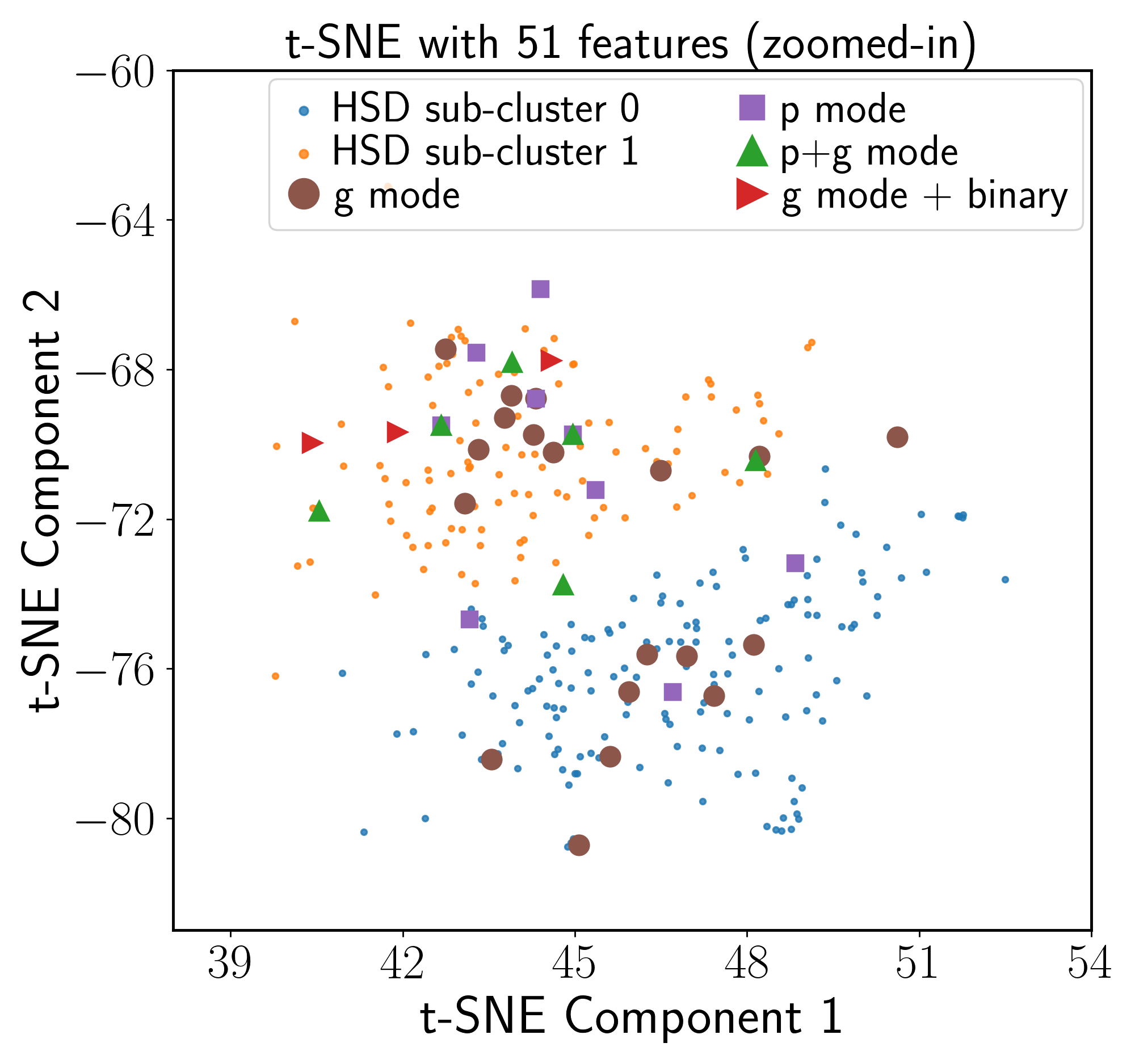} &
\includegraphics[width=0.42\linewidth] {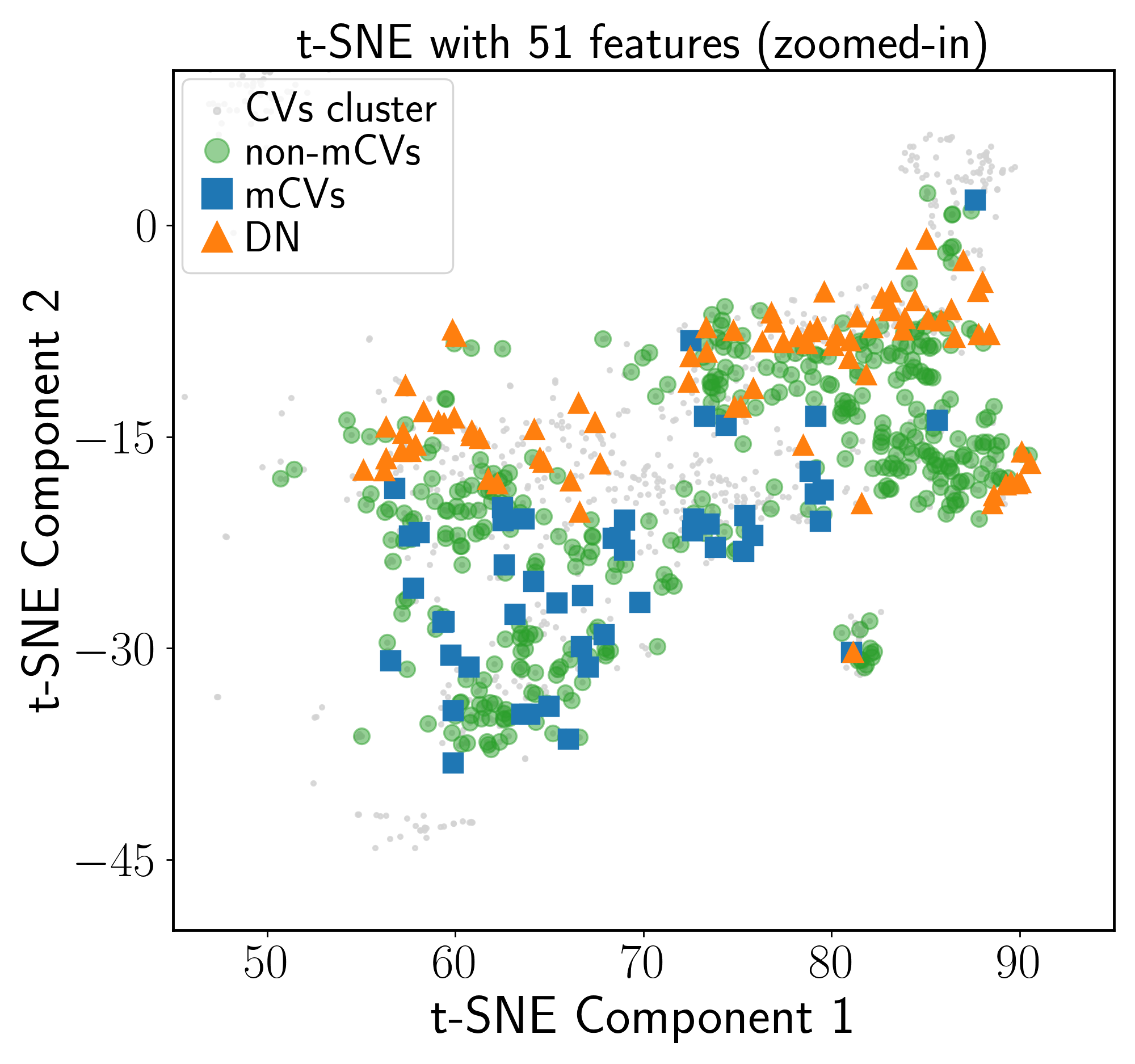}

\end{tabular}\caption{\small Close up view of the t-SNE embeddings for HSD (left panel) and CVs (right panel) clusters. Left panel: HSD sub-clusters 0 and 1 represent the cluster HSD in Fig.~\ref{fig:tsne_51_feat}–c, where p-mode hot subdwarfs were identified from \cite{Baran2024}, while the other modes (g, p+g, g mode + binary) were taken from \citep{Krzesinski2022}. Right panel: Magnetic CVs (mCVs), non-magnetic CVs (non-mCVs), and dwarf novae (DN) from \cite{Canbay2023} are shown.} 
    \label{fig:zoomedin_tsne}
\end{figure*}

Regarding the CVs cluster, the \texttt{stetson\_mag\_g\_fov} and the \texttt{mad\_mag\_g\_fov} contribute the most to the prediction of CVs, with Stetson variability index and median absolute deviation median values around 50 (against 3 for the full sample) and 0.28 mag (against 0.04 mag for the full sample), respectively. The values of these two parameters are consistent with the variability nature of CVs, where large-amplitude brightness variations are expected. Moreover, several variants of CVs were observed in the CVs cluster, including magnetic CVs (mCVs), non-magnetic CVs (non-mCVs), and dwarf novae (DN) from \cite{Canbay2023}. These sub classes are highlighted in the second panel of Fig.~\ref{fig:zoomedin_tsne}, where mCVs and DN tend to occupy two sub clusters. However, non-mCVs are ubiquitous in both sub clusters.

\subsubsection{Objects in the Gaia Andromeda photometric survey}\label{sec:GAPS}
The GAPS sample consists of an early release of epoch photometry of about 1.2 million sources centred on the Andromeda galaxy (M31), with a field radius of 5.5$\degree$ \citep{Evans2023}. Sources in the GAPS include objects within M31, or the Milky Way that happen to be in the line of sight. As introduced in Sect.~\ref{sec:results}, we found 2552 objects to be part of the GAPS survey. Since our initial target selection was limited to objects within 1 kpc, these objects are most likely Galactic objects. Their location in the t-SNE embeddings is shown in Fig.~\ref{subfig:b}, while the cluster with the most known GAPS objects is referred to as GAPS in Fig.~\ref{subfig:c}. By analysing their SHAP values, these objects are characterised by higher FAP values and low significance of variability (\texttt{log\_sigvar}) with median values of 0.2 and 0.35, respectively. These values could indicate weak detection of variability in the GAPS cluster. Since the GAPS survey also largely includes constant stars \citep{Evans2023}, such objects could contribute to the observed low variability significance in this cluster.

\begin{figure}
    \centering
    \begin{tabular}{c}
\includegraphics[width=0.92\linewidth]{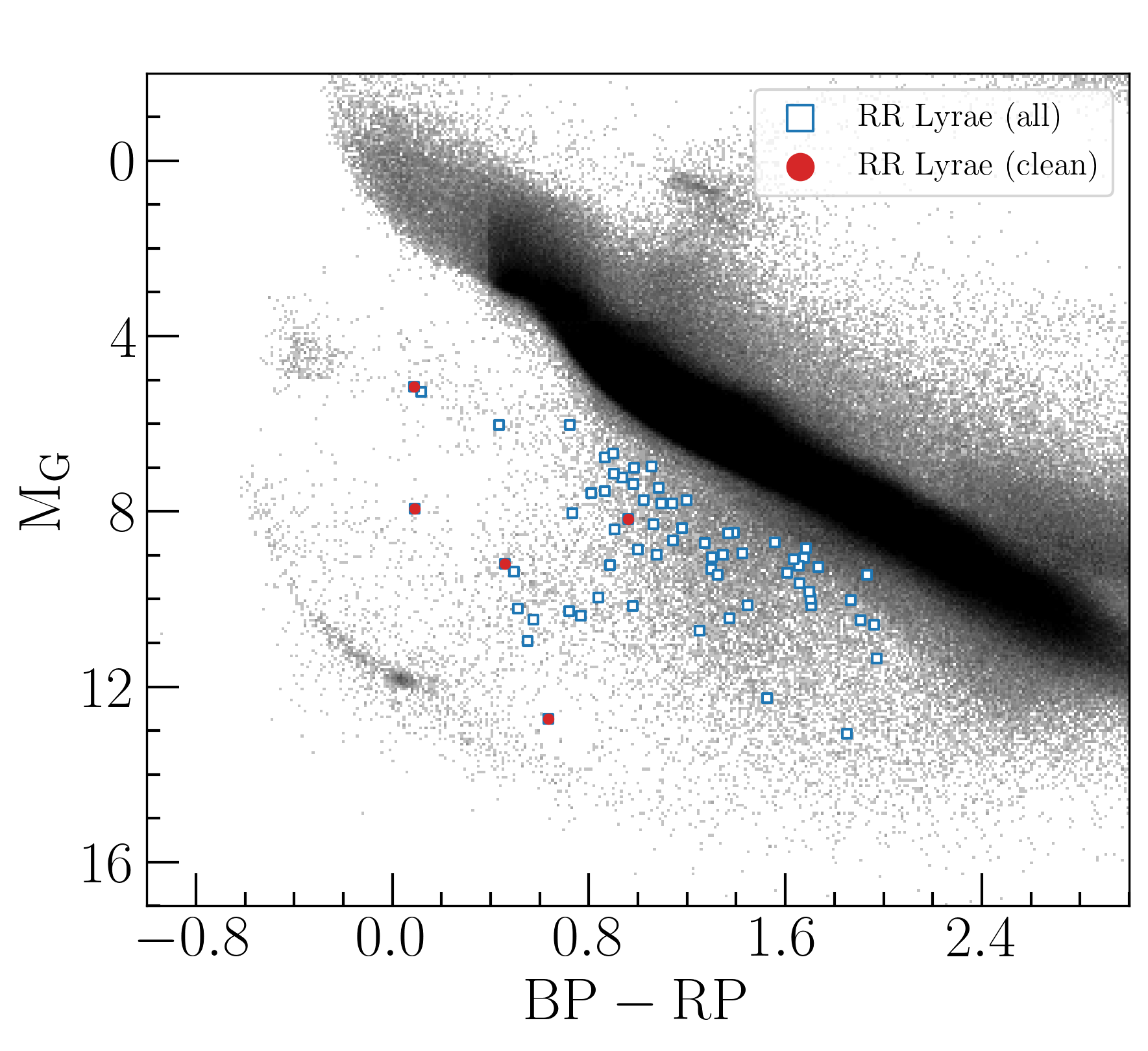}
\end{tabular}\caption{\small Colour-magnitude diagram of the 67 RR Lyrae stars identified from {\it Gaia} classification (blue squares) and the 5/67 objects (red circles) that met RR Lyrae selection criteria described in \cite{Iorio2021}. The grey background points representing all selected Gaia DR3 sources within 1 kpc.}
    \label{fig:cmd_rrlyrae}
\end{figure}

\begin{figure*}
    \centering
    \begin{tabular}{c}
   \includegraphics[width=0.92\linewidth]{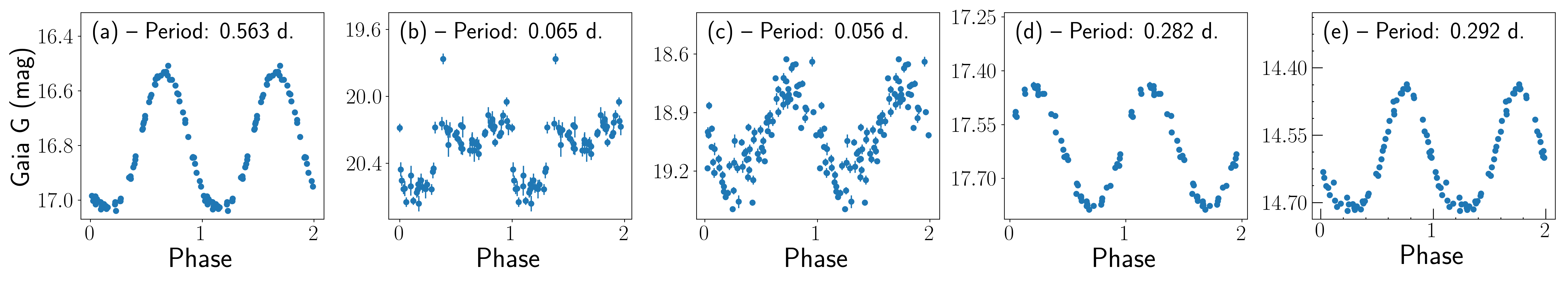}
    
    \end{tabular}\caption{\small {\it Gaia} light curves of five stars labelled as RR Lyrae passing the RR Lyrae selection criteria described in \cite{Iorio2021}. (a) Gaia DR3 378807525573579520, (b) Gaia DR3 5086653158769068928, (c) Gaia DR3 5281647899528664320, (d) Gaia DR3 5290302155549350272, (e) Gaia DR3 537040928284437632. }\label{fig:rrlyrae_lc}
\end{figure*}

\begin{figure*}
    \centering
    \begin{tabular}{c}
   \includegraphics[width=0.92\linewidth]{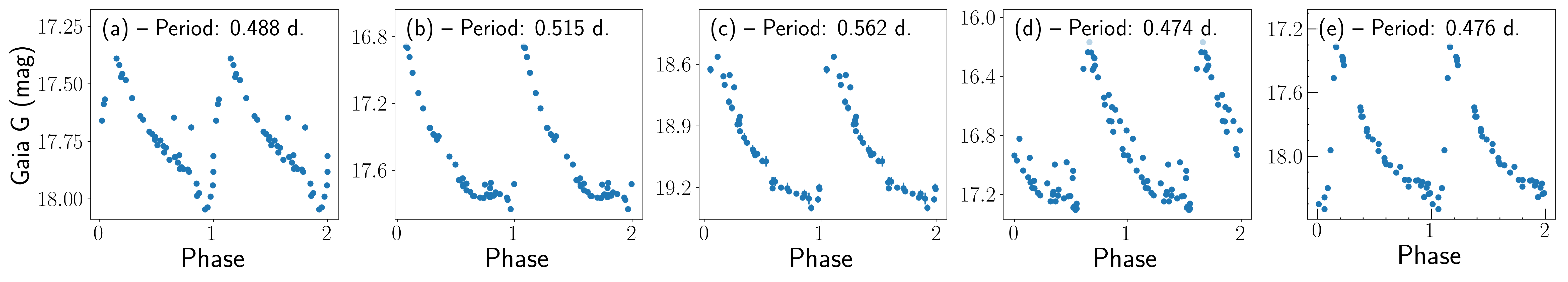}
    \end{tabular}\caption{\small {\it Gaia} light curves of five stars labelled as RR Lyrae that did not pass the RR Lyrae selection criteria described in \cite{Iorio2021}. (a) Gaia DR3 4107485483951148544, (b) Gaia DR3 4112601610223924480, (c) Gaia DR3 4122378020940824832, (d) Gaia DR3 4161748512969413888, (e) Gaia DR3 4268274211697325312. }\label{fig:rrlyrae_lc_ruwecut}
\end{figure*}
\subsection{RR Lyrae stars}\label{sec:rrlyrae}
As previously mentioned, 167 objects labelled as RR Lyrae stars from the {\it Gaia} SOS pipeline \citep{Clementini2023} were found in our sample. These are located in an unexpected location in the CMD – below the main sequence rather than above it (see Fig.~\ref{fig:cmd_rrlyrae}). More precisely, they fall within the ranges of {\it Gaia} $G$ absolute magnitude $\rm 5 < G_{abs} < 11$ and {\it Gaia} colour $\rm 0 < BP-RP < 2$, whereas RR Lyrae stars are typically expected to lie in the approximate range $\rm 0 < G_{abs} < 1 $ \citep{Garofalo2022} and $\rm 0 < BP-RP < 1$ (e.g. \citealt{Clementini2023, Lu2024}). Note that applying dust extinction and parallax zero-point offset corrections \citep{Garofalo2022} has only a minor effect on their positions in the CMD. To understand this misplacement, visual inspections of their light curves were first performed, revealing 67 objects with distinct RR Lyrae-like light curves, while the remaining 100 objects exhibit noisy light curves (e.g. Fig~\ref{fig:rrlyrae_noisy_lc}). Their derived periods and amplitudes from this work are consistent with that of RR Lyrae stars, with a median period and amplitude of 0.47 d and 0.26 mag, respectively. Among the 67 objects, 25 are also classified as RR Lyrae stars in the variable star index (VSX, \citealt{Watson2006}), excluding VSX classification from Gaia.

Secondly, the set of 67 objects with verified RR Lyrae-like light curves, amplitudes and periods were further examined by applying selection criteria described in \citealt{Iorio2021} to remove objects with unreliable astrometric measurements and contaminant sources in crowded fields. These criteria are based on the RUWE, the {\it Gaia} colour excess factor (\texttt{phot\_bp\_rp\_excess\_factor}), and the reddening $E(B-V)$ parameters. As a result of applying all three cuts, only 5 out of 67 objects remained, while the RUWE criterion alone (RUWE $<$ 1.2) retained 11 out of 67 objects. From their {\it Gaia} light curves alone (see Fig.~\ref{fig:rrlyrae_lc}), it is not obvious whether these five objects are genuine RR Lyrae stars. Three of them show regular, sinusoidal-like curves and could be eclipsing binary contaminants (e.g. WUMa-type variables), while the other two have periods shorter than expected for RR Lyrae stars and may instead be $\delta$~Scuti contaminants (e.g. Fig.~\ref{fig:rrlyrae_lc}, sub-panel c) or other types of variables.

On the other hand, Fig.~\ref{fig:rrlyrae_lc_ruwecut} shows a sample of five light curves of the objects that did not pass the RR Lyrae selection criteria. These objects exhibit unambiguous RR Lyrae-like (RRab) light curves.  However, since these stars were excluded by the three quality cuts, their estimated parallaxes may be systematically biased, and their uncertainties underestimated (e.g. \citealt{El-Badry2025}). One possible explanation is that these stars are part of unresolved binary systems. This has important implications for alternative RR Lyrae formation channels involving binary evolution (see, e.g. \citealt{Karczmarek2017, Bobrick2024}). To date, no RR Lyrae stars have been astrometrically confirmed as binaries \citep{Holl2023}. However, the upcoming {\it Gaia} data release DR4 will provide the opportunity to confirm or refute this scenario--both for the 67 RR Lyrae stars identified here and for the RR Lyrae population as a whole (Iorio et al., in prep.).
If, instead, the parallax measurements are not significantly affected by astrometric bias, their fainter absolute magnitudes ($G_{\mathrm{abs}} > 5~\mathrm{mag}$) may indicate that these are objects mimicking the RR Lyrae light curve, but with a different intrinsic nature or evolutionary pathway (e.g. \citealt{Pietrzynski2012}).

\section{Applying data quality cuts}\label{sec:ruwe_cut}
Inspired by the objects that appear as RR Lyrae, and since our initial targets were selected without applying any astrometric quality criteria, except for fractional parallax, we investigate the impact of applying a RUWE cut on the clustering results in this section. Although high RUWE values (e.g., RUWE $>$ 1.4) are potentially indicative of unresolved binary systems, other factors such as crowding and instrumental effects can also contribute to elevated RUWE values \citep{Castro-Ginard2024}. If, instead of our initial unconstrained selection, we apply a cut of RUWE $<$ 1.4, which corresponds to the upper limit of a sky-dependent RUWE threshold \citep{Castro-Ginard2024}, the number of objects in our sample drops to 6443.

This cut significantly affected the number of objects in nearly all clusters, with the exception of the CV and HSD clusters. Notably, the impact was strongest in the second short-timescale variables cluster (STS2), where the number of objects dropped from 1688 to just 34 after applying the RUWE cut. This is consistent with the SHAP value analysis shown in Fig.~\ref{fig:shap_values}, which indicates that RUWE is a dominant feature for classifying objects in this cluster.

For the cluster containing potential variables (EB1), 833 out of 1703 objects remained after applying the RUWE cut, of which 787 matched with visually confirmed bona fide variables. On the one hand, the RUWE cut improved the purity of the EB1 cluster from approximately 88\% (1497/1703 before the cut; see Sect.~\ref{sec:eclipsing_binaries}) to around 95\% (787/833 after the cut). On the other hand, it reduced the number of potential variables by nearly 50\%.

To evaluate the effect of applying the RUWE cut on the clustering results, the clustering steps described in Sect.~\ref{sec:tsne_optimisation} were repeated using the reduced dataset. Fig.~\ref{fig:tsne_51_feat}~(d–f) show the t-SNE embeddings generated using 46 features. In this new representation, the clusters corresponding to eclipsing binaries, white dwarfs, and hot subdwarfs appear more distinct than in the original embeddings as shown in Fig.~\ref{subfig:b} and Fig.~\ref{subfig:e}. This improvement could be due to the white dwarf and hot subdwarf classes being previously under-represented relative to the neighbouring eclipsing binary class. As most of the original clusters are now reduced in size due to the RUWE cut, their positions in the new t-SNE projection have shifted slightly, with the short-timescale variable cluster showing the most notable change. Additionally, some contamination is visible across clusters in the new t-SNE embeddings shown in Fig.~\ref{subfig:f}, where the original cluster labels from Fig.~\ref{subfig:c} are used. This is because data points that previously had neighbours from the removed data may now be drawn to different nearby points and consequently shifting their location. These observations highlight the sensitivity of t-SNE to sample distribution and emphasise the critical role of sampling in shaping the resulting low-dimensional structures, potentially revealing or obscuring important patterns in the data \citep{Vandermaaten2008,Policar2021}. 

\section{Conclusion and future prospects}\label{sec:conclusion}
The unsupervised ML framework developed in Paper I was extended in this work to classify {\it Gaia} light curves for objects located between the main sequence and the white dwarf sequence. Instead of the 1576 pre-selected targets under scrutiny in Paper I, the current analysis was based on 13\,405 objects with at least 25 observations in the {\it Gaia} G band located in a much wider region of the Gaia CMD. Following the feature extraction and selection procedures outlined in Paper I, 51 features were selected and used as the basis for the unsupervised clustering using t-SNE. For data treated here, these 51 features yielded better cluster separation in the t-SNE embeddings than the 27 features selected in Paper I.

To assess the integrity of the clusters observed in the t-SNE embeddings and to gain insights into the nature of each cluster, objects with known classifications were overplotted onto the embeddings. This cross-matching helped identify the number of distinct clusters in the t-SNE representation, revealing 10 clusters and sub-clusters. This number was used as the input for a Gaussian mixture model to assign objects to their corresponding clusters. The 10 clusters were further examined using SHAP values, which highlighted the most important features characterising each cluster. In addition, the clustering analysis was repeated on a reduced dataset of 6443 objects to assess the impact of applying a RUWE cut on the t-SNE clustering and classification results. 

Two distinct clusters for known hot subdwarfs and CVs were detected in the t-SNE embeddings, which is consistent with the findings in Paper I. In addition, this analysis helped the identification of a cluster of objects (EB1) with pure photometric variability, including eclipsing binaries, hot subdwarfs, and white dwarfs. Key features for identifying this cluster include the \texttt{p95\_100} and \texttt{n05} parameters introduced in Paper I. Clusters associated with spurious variability and in crowded fields were also detected (STS1, STS2, GAPS, EB3); these objects typically display slightly different RUWE and FAP distributions.

As for the impact of RUWE filtering on the classification, the results indicate that it can effectively remove spurious or noisy data, revealing under-represented classes, such as white dwarfs and hot subdwarfs. While this cut eliminates many spurious variables, it also discards a significant fraction of potential variables, particularly eclipsing binaries. This is expected, as eclipsing binaries often exhibit high RUWE values, although other factors may also contribute to elevated RUWE. The decision to apply a RUWE cut should therefore be guided by the specific object types of interest. For instance, in the case of hot subdwarfs, a relaxed threshold of RUWE$<$7 has been applied by \cite{Dawson2024} to avoid excluding promising candidates.

This work also led to the identification of 67 objects that were classified as RR Lyrae stars in the {\it Gaia} SOS pipeline, which exhibit all typical characteristics of RR Lyrae stars, yet are located in an unusual place in the CMD. Analysis of their astrometric parameters and light curves proposed three possible explanations: either their positions in the CMD result from poor astrometric measurements; they represent a different evolutionary channel for RR Lyrae stars; or they represent an evolutionary channel for objects that display features very similar to classical RR Lyrae stars.

The findings of this study suggest several implications. First, the proposed unsupervised ML framework is scalable to large datasets with rich variety of stellar populations. Second, this approach is not limited to detecting photometric variability; it also aids in identifying instrumental effects and anomalies, which could facilitate faster analysis of large-scale datasets. Third, the results of this study present the possibility of identifying sub classes or intrinsic properties of a given stellar population, such as pulsation modes in hot subdwarfs, based only on statistical parameters. This is particularly valuable for increasing the detection of under-represented classes in population studies. We note that the Gaia classifications and literature-based class labels from literature shown in Fig.~\ref{subfig:b} are not used as a training set or as ground truth in our analysis. Our embedding (Fig.~\ref{subfig:a}) is derived in a fully unsupervised manner from light-curve features. The Gaia labels are included only as an external reference to illustrate how broadly defined variability classes are distributed in the embedding. While these classes are known to be imperfect and in some cases biased (see e.g. \citealt{Rimoldini2023,Gavras2023}), they remain useful to explore specific Gaia-defined categories in this representation. Since the clustering algorithms used here were designed to embed new data points into existing t-SNE embeddings \citep{Policar2021}, the framework can accommodate new datasets without the need for retraining. Further research may explore the performance of the proposed ML approach on data from other observations, notably those from ground-based telescopes, such as the BlackGEM telescopes \citep{Groot2024}.  

\section{Data availability}\label{sec:data_availability}
The complete version of Table~\ref{tab:target_sample_list}, containing the classifications of the 13,405 targets, will be made available in electronic form at the CDS via anonymous ftp to cdsarc.u-strasbg.fr (130.79.128.5) or via \url{http://cdsweb.u-strasbg.fr/cgi-bin/qcat?J/A+A/}.

\begin{acknowledgements}
C.J.~acknowledges funding from the Royal Society through the Newton International Fellowship funding scheme (project No. NIF$\backslash$R1$\backslash$242552). This research was supported by Deutsche Forschungsgemeinschaft  (DFG, German Research Foundation) under Germany’s Excellence Strategy - EXC 2121 "Quantum Universe" - 390833306. Co-funded by the European Union (ERC, CompactBINARIES, 101078773). Views and opinions expressed are however those of the author(s) only and do not necessarily reflect those of the European Union or the European Research Council. Neither the European Union nor the granting authority can be held responsible for them. The research leading to these results has received funding from the Research Foundation Flanders (FWO) under grant agreement
G0A2917N (BlackGEM), as well as from the BELgian federal Science Policy Office (BELSPO) through PRODEX grants for {\it Gaia} data exploitation.
This work has made use of data from the European Space Agency (ESA) mission {\it Gaia} (https://www.cosmos.esa.int/Gaia), processed by the {\it Gaia} Data Processing and Analysis Consortium (DPAC, https://www.cosmos.esa.int/web/Gaia/dpac/consortium). Funding for the DPAC has been provided by national institutions, in particular the institutions participating in the {\it Gaia} Multilateral Agreement. PJG is supported by NRF SARChI grant 111692. 

\end{acknowledgements}

\bibliographystyle{aa}
\bibliography{main}
\begin{appendix}
\onecolumn
\section{Additional material}\label{appendix:A}

\subsection{Gaia ADQL query}
\begin{lstlisting}[language=SQL, basicstyle=\ttfamily\small, frame=single,showstringspaces=false,numbers=left,breaklines=true]
SELECT source_id, ra, dec, parallax, parallax_error, phot_g_mean_mag, bp_rp, parallax_over_error, num_selected_g_fov FROM gaiadr3.gaia_source
INNER JOIN gaiadr3.vari_summary AS var USING (source_id)
WHERE
parallax > 1 AND
parallax_over_error > 5 AND 
has_epoch_photometry = 'TRUE' AND num_selected_g_fov > 24

\end{lstlisting}
\subsection{Gaia summary statistic table query}
\begin{lstlisting}[language=SQL, basicstyle=\ttfamily\small, frame=single,showstringspaces=false,numbers=left,breaklines=true]
SELECT target.*, gaia.* FROM gaiadr3.vari_summary AS gaia, user_username.table1 AS target WHERE target.source_id IN (gaia.source_id)

\end{lstlisting}

\begin{figure*}[hb]
    \centering
    \begin{tabular}{c}
   \includegraphics[width=0.25\linewidth]{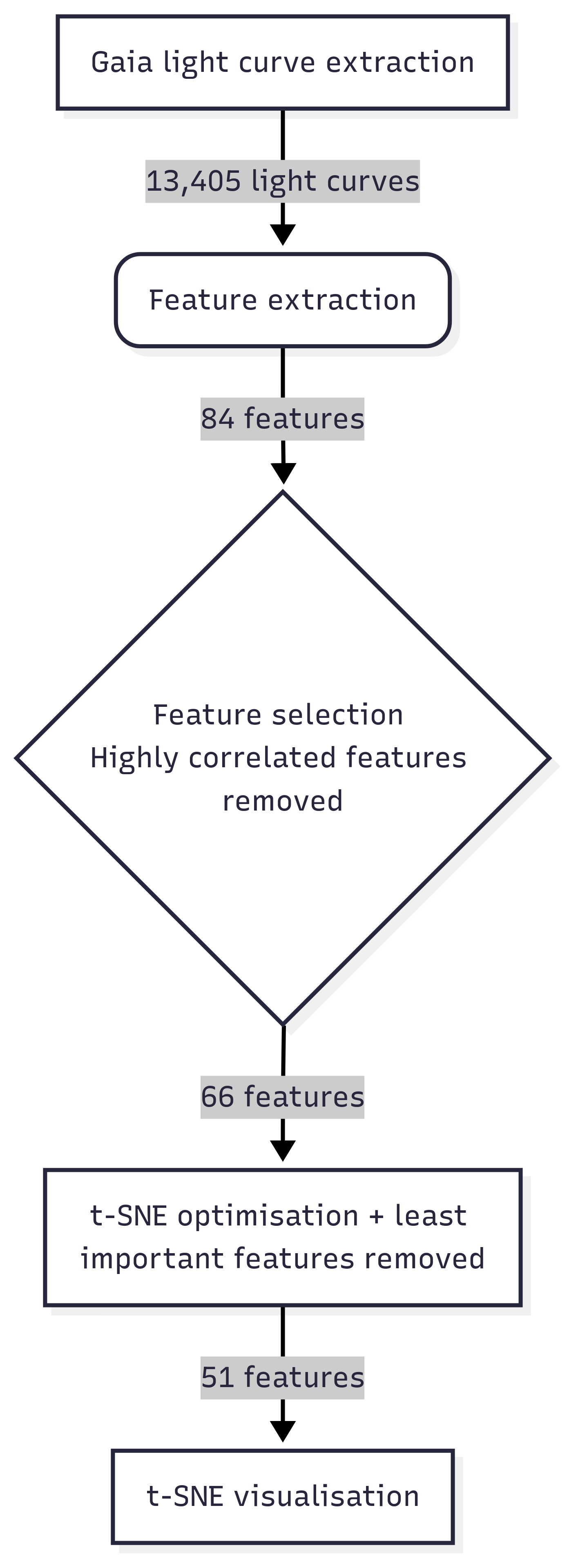}
    
    \end{tabular}\caption{Flowchart summarising the dimensionality reduction steps using t-SNE.}\label{fig:tsne_flowchart}
\end{figure*}

\begin{figure*}[!hb]
    \centering
    \begin{tabular}{c}
   \includegraphics[width=0.75\linewidth]{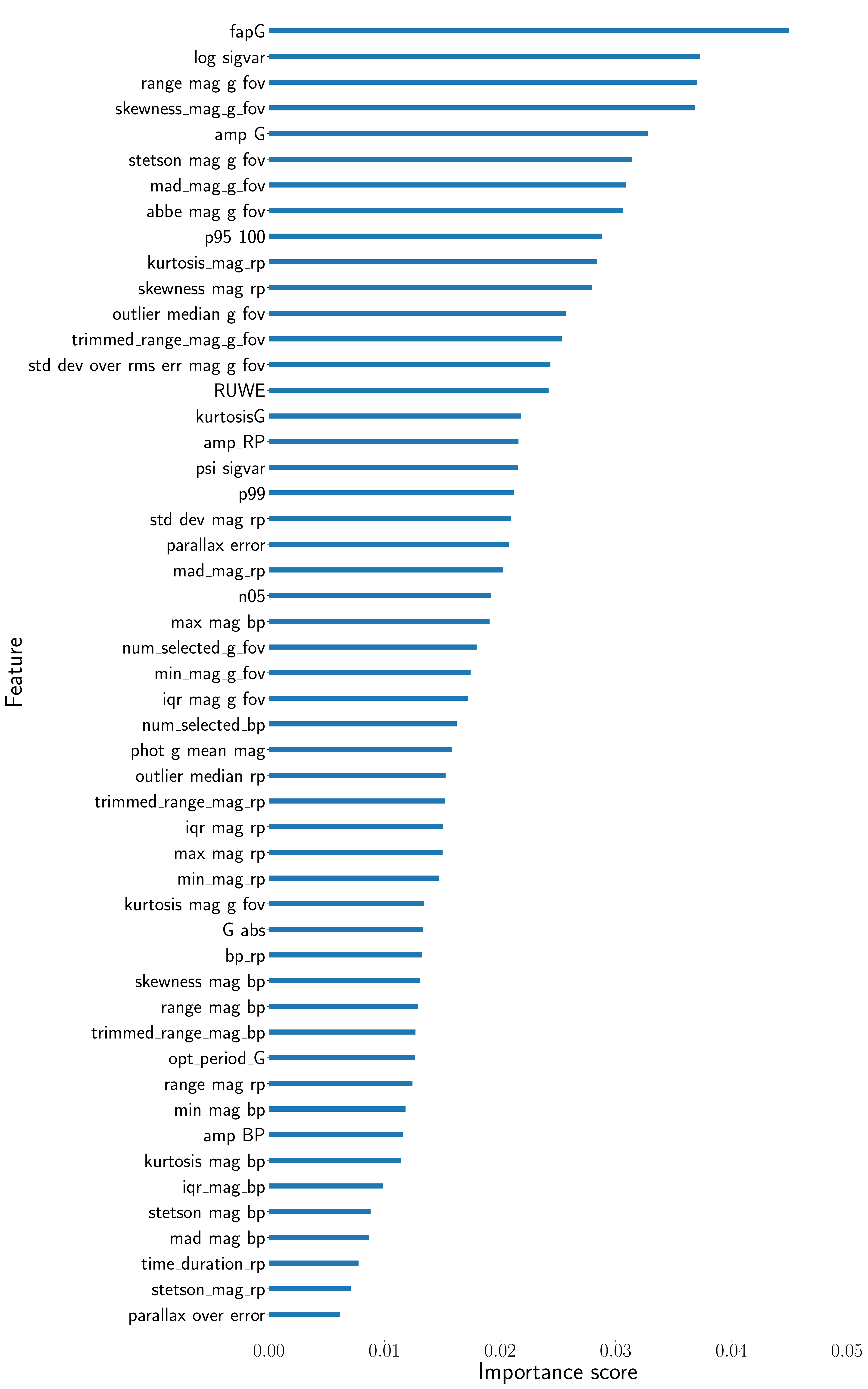}
    \end{tabular}\caption{Random Forest feature importance scores for the selected 51 features.}\label{fig:feature_importance_score}
\end{figure*}

\begin{figure*}
    \centering
    \begin{tabular}{c}
   \includegraphics[width=0.5\linewidth]{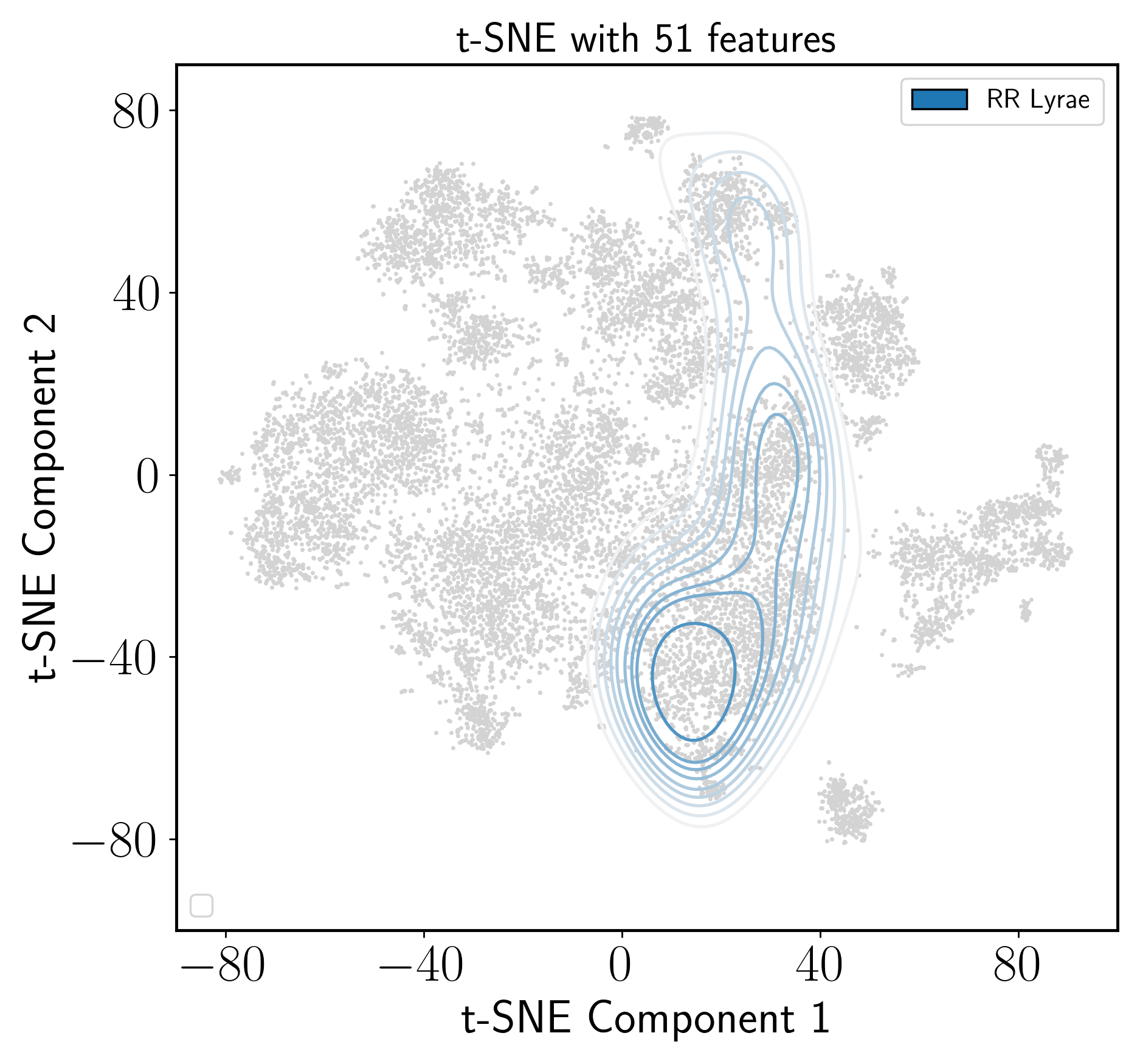}
    
    \end{tabular}\caption{t-SNE embeddings depicting the distribution of RR Lyrae stars classified by {\it Gaia}. }\label{fig:rrlyrae_tsne}
\end{figure*}
\begin{figure*}
    \centering
    \begin{tabular}{c}
   \includegraphics[width=0.95\linewidth]{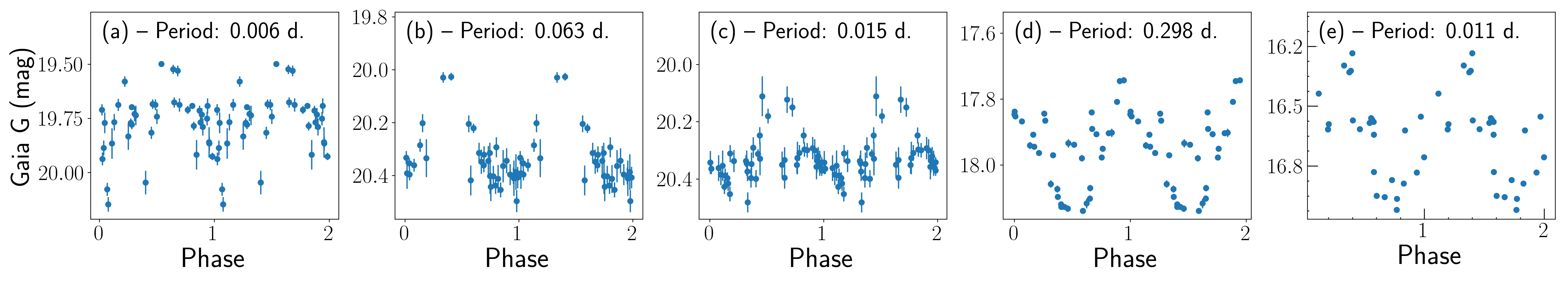}
    \end{tabular}\caption{{\it Gaia} light curves of five stars labelled as RR Lyrae that did not pass the RR Lyrae selection criteria described in \cite{Iorio2021}, which exhibit noisy light curves or spurious variability. (a) Gaia DR3 4325299252697361920, (b) Gaia DR3 5850070779543826944, (c) Gaia DR3 6056717633367527552, (d) Gaia DR3 4056072560643550336, (e) Gaia DR3 4042776681999969152. }\label{fig:rrlyrae_noisy_lc}
\end{figure*}

\begin{table*}
    \centering
    \caption{Description of the object type labels in Fig.~\ref{tab:nobj_per_cluster}.}\label{tab:obj_definition}
    \begin{tabular}{ll}
    Object type  & description \\ \hline
    {\small \tt EB\_G} & Eclipsing binary \\
    {\small \tt short\_TS\_G} & Short timescale \\
    {\small \tt Rot\_SOS\_G} & Rotational modulation (Gaia SOS pipeline classification)\\
    {\small \tt Rot\_ML\_G} & Rotational modulation (Gaia machine learning classification)\\
    {\small \tt GAPS\_G} & M31 field \\
    {\small \tt CV\_lit} & Cataclysmic variables in \cite{Canbay2023} catalogue\\
    {\small \tt Hsd\_C0} & Hot subdwarfs in Paper I's cluster 0\\
    {\small \tt Hsd\_C1} & Hot subdwarfs in Paper I's cluster 1 \\
    {\small \tt WD\_SB} & White dwarfs from SIMBAD\\
    RR Lyrae& RR Lyrae stars (Gaia SOS classification)\\
    {\small \tt Short\_TS\_GaiaSOS} & Short timescale (Gaia SOS classification)\\
    {\small \tt sdB\_pulsators} & Pulsating hot subdwarf B stars \\\hline
    \end{tabular}
\tablefoot{ 
The {\tt G} annotation denotes object classes from Gaia DR3 classifications. The {\tt SOS} and {\tt ML} annotations in the object type column indicate objects classified by the Gaia SOS and ML pipelines, respectively.} 
\end{table*}

\begin{figure*}
    \centering
    \begin{tabular}{c}
   \includegraphics[width=0.65\linewidth]{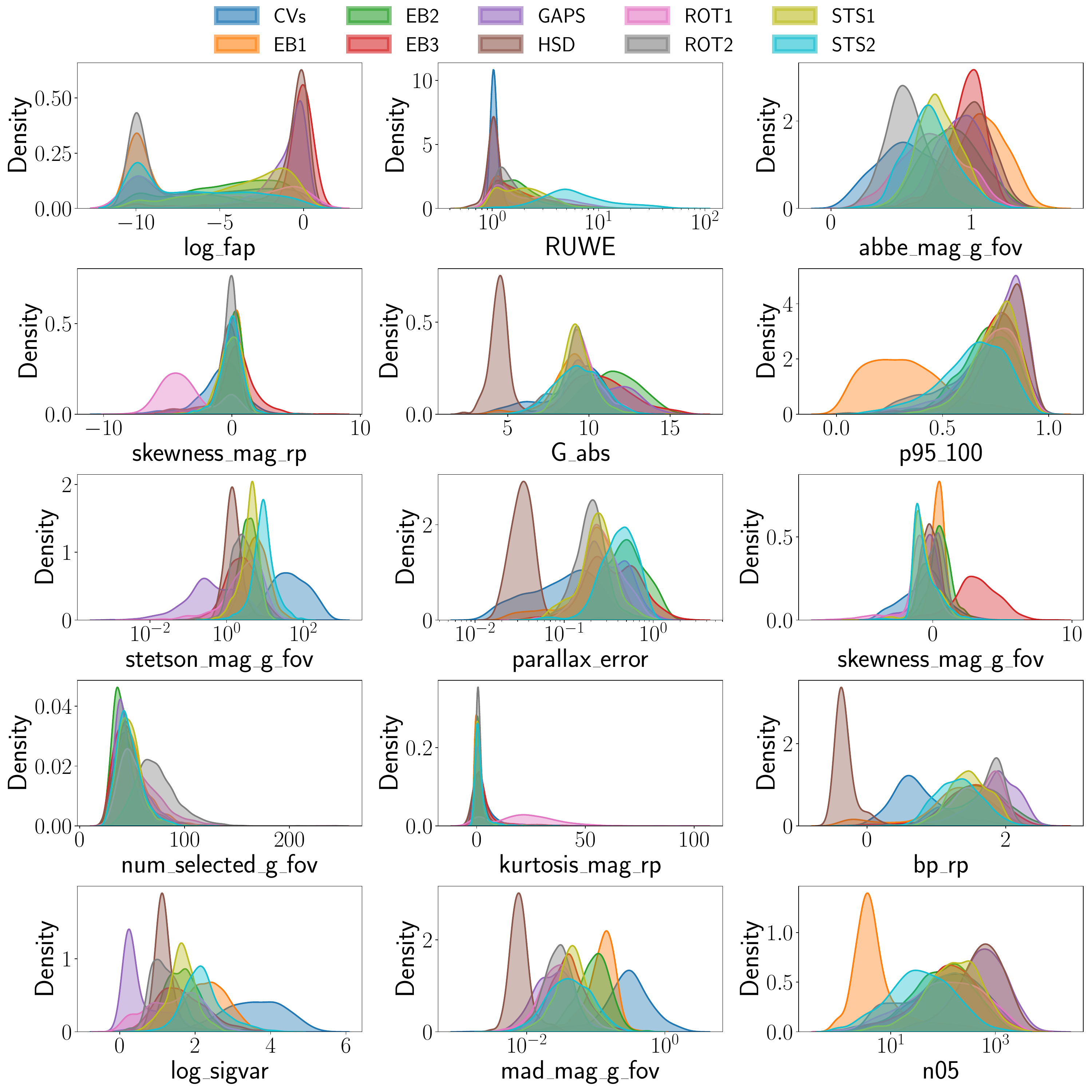}
    
    \end{tabular}\caption{Kernel density estimate (kde) plots for features with high importance scores from SHAP values. }\label{fig:kde_plots_features}
\end{figure*}
\begin{figure*}
    \centering
    \begin{tabular}{c}
   \includegraphics[width=0.90\linewidth]{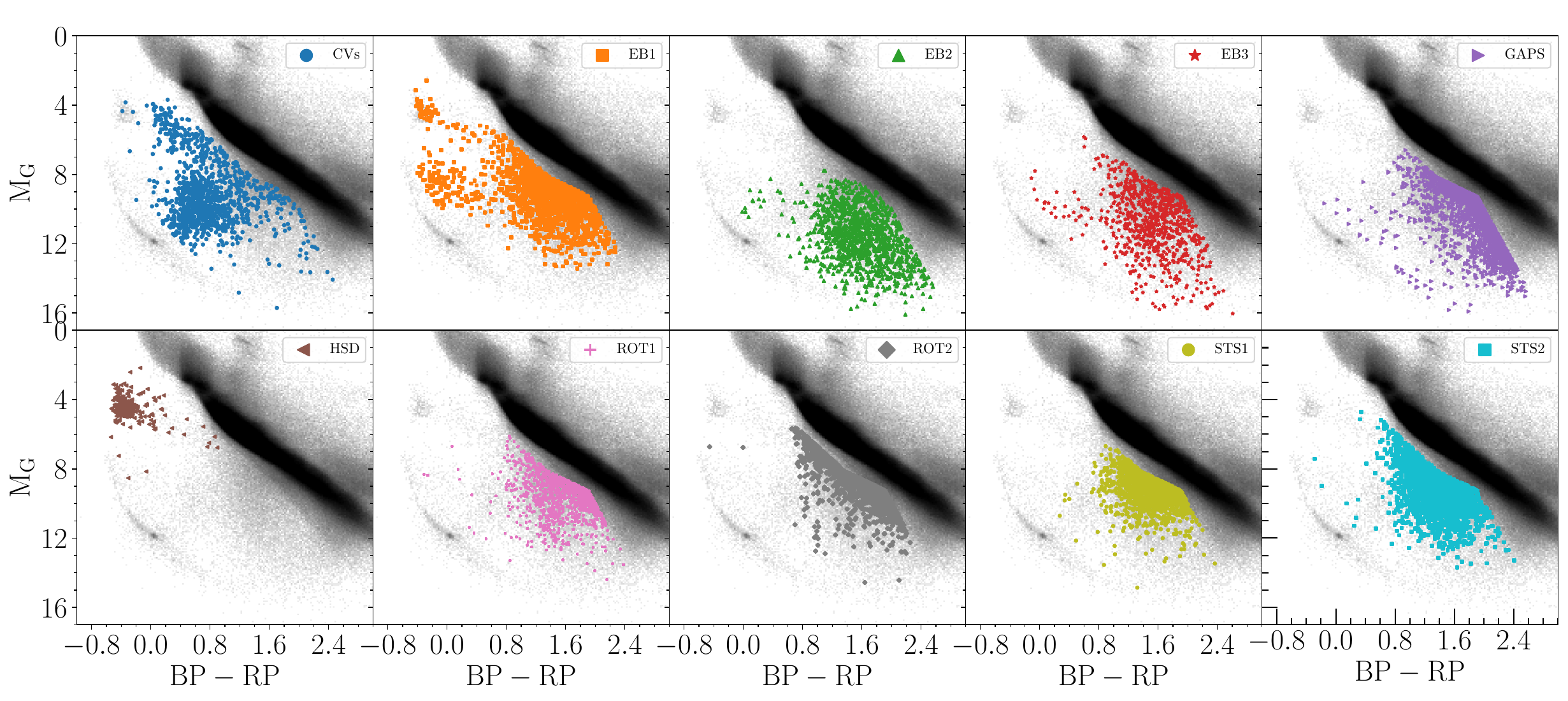}
    
    \end{tabular}\caption{Colour-magnitude diagram of each cluster shown in Fig.~\ref{subfig:c} }\label{fig:cmd_each_cluster}
\end{figure*}

\begin{figure*}
    \centering
    \begin{tabular}{c}
   \includegraphics[width=0.90\linewidth]{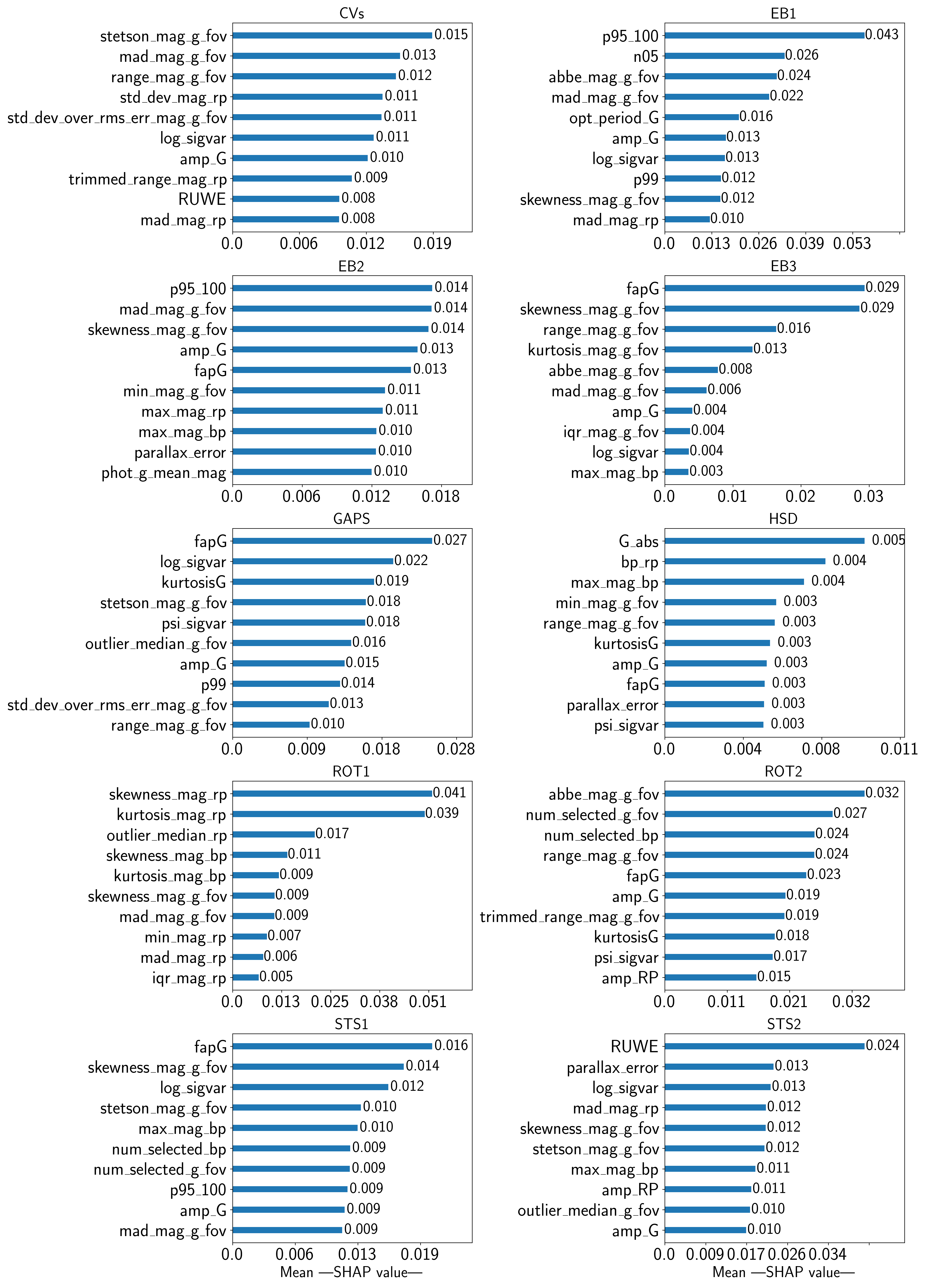}
    
    \end{tabular}\caption{Top 10 most important features per cluster. }\label{fig:top10_features}
\end{figure*}

\begin{landscape}
\begin{table}
    \centering
    \setlength{\tabcolsep}{3pt}
    \caption{List of 13\,405 targets with their stellar and variability classifications, along with their t-SNE embeddings.}
    \label{tab:target_sample_list}
    \begin{tabular}{cccccccccccccc}
    \toprule
GaiaDR3	&	RA	&	DEC	&	G 	&	G abs	&	BP-RP	&	Period (G)	&	tsne comp1	&	tsne comp2	&	Cluster	&	Gaia SOS	&	Gaia ML	&	Lit.	&	Lit. ref\\
	&	(deg)	&	(deg)	& (mag)	& (mag)	&		&	(day)	&	&		&  name	& class	&	class	&	class	&		\\ \midrule 
1250382352732030592	&	209.20260	&	21.08615	&	12.11	&	2.17	&	-0.14	&	0.00312	&	49.31339724	&	-77.40044402	&	HSD	&	Short TS	&	-	&	Hot sd 	&	2022A\&A...662A..40C	\\
6678479845256836224	&	313.33771	&	-40.10817	&	12.09	&	2.42	&	-0.28	&	0.00715	&	44.13395004	&	-66.90342013	&	HSD	&	Short TS	&	-	&	Hot sd 	&	2022A\&A...662A..40C	\\
5362804330246457344	&	163.66888	&	-48.78408	&	12.14	&	2.58	&	-0.28	&	0.35711	&	15.98243033	&	-70.19340873	&	EB1	&	Short TS	&	-	&	Hot sd	&	2025A\&A...693A.268R	\\
1561116845686660352	&	208.81621	&	53.57852	&	12.99	&	3.07	&	-0.36	&	0.23186	&	44.71416154	&	-68.38341803	&	HSD	&	Short TS	&	-	&	Hot sd 	&	2022A\&A...662A..40C	\\
837007590331263360	&	161.72113	&	51.90995	&	12.49	&	3.12	&	-0.50	&	0.00368	&	44.9649494	&	-67.86354242	&	HSD	&	Short TS	&	-	&	Hot sd 	&	2022A\&A...662A..40C	\\
4962221462214625920	&	25.78132	&	-38.55447	&	12.94	&	3.14	&	-0.50	&	0.01236	&	48.21751721	&	-68.91068293	&	HSD	&	Short TS	&	-	&	Hot sd &	2025A\&A...693A.268R	\\
5661504084315014656	&	143.70102	&	-25.21241	&	13.04	&	3.15	&	-0.42	&	0.14290	&	18.98356475	&	-70.97223765	&	EB1	&	Short TS	&	-	&	Hot sd 	&	2022A\&A...662A..40C	\\
4299431347569705216	&	303.40667	&	9.46707	&	12.38	&	3.15	&	-0.40	&	0.00364	&	43.27705812	&	-67.53818221	&	HSD	&	Short TS	&	-	&	Hot sd 	&	2022A\&A...662A..40C	\\
3176695152292552704	&	63.33094	&	-13.68413	&	12.48	&	3.16	&	-0.47	&	0.01156	&	42.86458933	&	-67.59454435	&	HSD	&	-	&	sdB	&	Hot sd 	&	2022A\&A...662A..40C	\\
1553487166999658112	&	201.00285	&	49.37568	&	12.41	&	3.19	&	-0.33	&	0.00282	&	44.97884981	&	-67.85417212	&	HSD	&	Short TS	&	WD	&	Hot sd 	&	2022A\&A...662A..40C\\
\vdots&\vdots&\vdots&\vdots&\vdots&\vdots&\vdots&\vdots&\vdots&\vdots&\vdots&\vdots&\vdots&\vdots\\

\\\hline
    \end{tabular}
\tablefoot{The full table will be made available at the CDS.}
\end{table}

\end{landscape}

\end{appendix}
\end{document}